\documentclass[12pt,draftcls,onecolumn]{IEEEtran}
\usepackage{dsfont}

\usepackage{color}
\usepackage{setspace}
\usepackage{clrscode}
\usepackage{amsthm}
\usepackage{pict2e}
\usepackage{amsmath}
\usepackage{amssymb}
\usepackage{graphicx}
\usepackage{bbold}
\usepackage{algorithm}
\usepackage{algorithmic}
\usepackage{caption}
\usepackage{subcaption}
\usepackage{mathabx}
\usepackage{stfloats}
\usepackage[ps2pdf,
bookmarks=false,
bookmarksnumbered=false, % true means bookmarks in
 left window are numbered
bookmarksopen=false, % true means only level 1
 are displayed.
colorlinks=false]{}

\DeclareMathOperator*{\argmax}{argmax}
\DeclareMathOperator*{\argmin}{argmin}

\newcommand{\cL}{\mathcal{L}}
\newcommand{\Fbar}{{\overline{F}}}
\newcommand{\prob}{\mathbb {P}}

\newcommand{\los}{{\text{{LOS}}}}
\newcommand{\nlos}{{\text{{NLOS}}}}
\newcommand{\sinr}{{\text{{SINR}}}}
\newcommand{\suc}{{\text{{suc}}}}

\newcommand{\E}{\mathbb{E}}
\newcommand{\indictor}{\mathbb{1}}
\newcommand{\ahat}{\hat{a}}
\newcommand{\fhat}{\hat{f}}
\newcommand{\Fhat}{\hat{F}}
\newcommand{\xbf}{\textbf{x}}
\newcommand{\ybf}{\textbf{y}}
\newcommand{\dsp}{{\rm d}}
\newcommand{\Ksp}{\mathcal{K}}

\newtheorem{Theo}{Theorem}
\newtheorem{Lem}{Lemma}
\newtheorem{Rem}{Remark}

\newtheorem{Def}{Definition}

\allowdisplaybreaks[2]
\begin{document}
\title{Coverage Analysis for Energy-Harvesting UAV-assisted mmWave Cellular Networks}

%\author{Xueyuan Wang, M. Cenk Gursoy
%	\\Department of Electrical Engineering and Computer Science,
%	Syracuse University, Syracuse, NY 13244
%	\\Email: xwang173@syr.edu, mcgursoy@syr.edu}
\author{\IEEEauthorblockN{Xueyuan Wang and M. Cenk Gursoy}
	\thanks{The authors are with the Department of Electrical
		Engineering and Computer Science, Syracuse University, Syracuse, NY, 13244
		(e-mail: xwang173@syr.edu,  mcgursoy@syr.edu).}
	\thanks{Manuscript received May 27, 2019; revised September 1, 2019.}}

\maketitle

\begin{abstract}
	In this paper, we jointly consider the downlink simultaneous wireless information and power transfer (SWIPT) and uplink information transmission in  unmanned aerial vehicle (UAV)-assisted millimeter wave (mmWave) cellular networks, in which the user equipment (UE) locations are modeled using Poisson cluster processes (e.g., Thomas cluster processes or Mat\'ern cluster processes). Distinguishing features of mmWave communications, such as different path loss models for line-of-sight (LOS) and non-LOS (NLOS) links and directional transmissions are taken into account. In the downlink phase, the association probability, and energy coverages of different tier UAVs and ground base stations (GBSs) are investigated. Moreover, we define a successful transmission probability to jointly present the energy and signal-to-interference-plus-noise ratio (SINR) coverages and provide general expressions. In the uplink phase, we consider the scenario that each UAV receives information from its own cluster member UEs. We determine the Laplace transform of the interference components and characterize the uplink SINR coverage. In addition, we formulate the average uplink throughput, with the goal to identify the optimal time division multiplexing between the donwlink and uplink phases.  Through numerical results we investigate the impact of key system parameters on the performance. We show that the network performance is improved when the cluster size becomes smaller. In addition, we analyze the optimal height of UAVs, optimal power splitting value and optimal time division multiplexing that maximizes the network performance.
\end{abstract}

\thispagestyle{empty}
\begin{spacing}{1.3}

\section{Introduction}
As an emerging technology, the Internet of Things (IoT) is expected to offer promising solutions to transform the operation and role of many existing industrial systems such as transportation systems and manufacturing systems \cite{IoT_LXu}. Expected to be commercially available in early 2020s, the fifth generation (5G) enabled IoT will connect massive number of IoT devices \cite{IoT_SHLi} \cite{EH_JTang}. In  certain applications, IoT sensors are low-power devices. In such cases, radio frequency (RF) energy harvesting is thus considered as an appealing solution to provide perpetual and cost-effective energy supply to power-constrained wireless devices \cite{UAV_EH_YWu}\cite{UAV_EH_JXu}\cite{UAV_EH_LXie}, and it is anticipated to lead to numerous applications in future wireless IoT networks \cite{EH_XLu}.
In conventional wireless power transfer (WPT) systems, energy transmitters are deployed at fixed locations, and therefore due to the RF signal propagation over potentially long distances, such systems can suffer from low end-to-end energy transfer efficiency \cite{UAV_EH_YWu}.
In general, RF WPT is considered in the context of two key application scenarios, namely simultaneous wireless information and power transfer (SWIPT) and wireless powered communication networks (WPCNs). SWIPT explores the dual use of microwave signals to achieve WPT and wireless information transfer (WIT) both in downlink direction \cite{EH_ZHu}, while downlink WPT and uplink WIT are performed in WPCNs \cite{EH_SBi}.

Unmanned aerial vehicles (UAVs) have emerged as key enablers of seamless wireless connectivity in diverse scenarios such as large-scale temporary events, military operations and disaster scenarios, and of capacity enhancement in the occasional demand of super dense base stations (BSs), and UAVs are anticipated to be part of future generation wireless networks \cite{UAV_YZeng}\cite{UAV_PSharma}\cite{UAV_LZhou}\cite{UAV_CLiu}.
More specifically, in order to take advantage of flexible deployment opportunities \cite{UAV_LZhou}, and high possibility of line-of-sight (LoS) connections with a ground user equipment (UE) \cite{UAV_AHourani}, BSs can be mounted on UAVs to support wireless connectivity and improve the performance of cellular networks \cite{UAV_CLiu}.
The flexibility of UAV BSs allows them to adapt their locations to the demand of UEs \cite{UAV_CLiu}, leading also to a new UAV-assisted WPT architecture.
Moreover, such UAV-assisted communication systems are drawing attention from the IoT community as well \cite{UAV_BGalkin}.  These potential benefits and improvements motivate further studies on performance of the UAV-assisted cellular networks.

The system level analysis of a network strongly depends on the deployment of the BSs and the UEs. In most recent WPT UAV-assisted network analysis, BSs and UEs locations are modeled as independent Poisson point processes (PPPs). However, in practice UEs are expected to be more densely distributed in the areas where the UAV-BSs are deployed, e.g., in large temporary events, disaster areas. In an RF-powered IoT network, UAV are deployed to collect data from an area where there is a concentration of IoT UEs or there exists macro BS coverage deadzones. This naturally couples the locations of the UEs and the UAV locations. The third generation partnership project (3GPP) has considered the clustered configurations in which locations of the UE and small-cell BSs (UAVs) are coupled, in addition to the uniformly distributed UEs \cite{cluster_CSahaPCP_conf}.
Therefore, Poisson cluster processes (PCPs) can provide  accurate models for the UE distribution in a UAV-assisted cellular network, in which the UEs are clustered around the projection of the UAVs on the ground.

\subsection{Related Studies}
Recently, UAV-assisted WPT systems have been intensively studied in the literature.
For instance, the authors in \cite{UAV_EH_YWu} considered a system where a UAV was dispatched to deliver wireless energy to charge two energy receivers (ERs) on the ground. The energy received by the two ERs was maximized by jointly optimizing the altitude, trajectory, and transmit beamwidth of the UAV.
\cite{UAV_EH_JXu} considered a more general scenario with a set of ERs, where the goal was to maximize the amount of energy transferred to all ERs by trajectory control.
In \cite{UAV_EH_LXie}, a WPCN scenario was addressed, where one mobile UAV could charge multiple ground UEs in downlink, and the UEs use the harvested RF energy to send information to the UAV in uplink.

The system-level analysis of UAV-assisted networks has also attracted much attention in recent literature.
For instance, references \cite{UAV_LZhou}, \cite{UAV_CLiu} and \cite{UAV_BGalkin}  considered a two dimensional (2D) PPP UAV-assisted cellular network, where UAVs were distributed according to a PPP at the same height in the air. In \cite{UAV_LZhou}, the downlink coverage probability was explored, as well as the influence of UAV height and density.
In \cite{UAV_CLiu}, different path loss models for high-altitude, low-altitude and ultra-low-altitude models were discussed. In addition to the coverage probability, the area spectral efficiency was investigated.
 The model  in \cite{UAV_BGalkin} also took into account the system parameters such as building density and UAV antenna beamwidth.
Besides the 2D PPP distributed UAV-assisted cellular networks, the authors in \cite{UAV_PSharma} considered a network in which a serving UAV was assumed to be located at fixed altitude, while a given number of interfering UAVs were assumed to have three dimensional (3D) mobility based on the mixed random waypoint mobility.
Moreover, \cite{UAV_VChetlur} considered a finite UAV network which was modeled as a uniform binomial point process (BPP).
Several limiting cases were discussed, including the no fading case and the dominant interferer based case.
% In addition, A $K$-tier uplink cellular networks with RF energy harvesting was considered in \cite{EH_ASakr}. BSs and UEs are modeled as two independent PPPs, and tools from queueing theory is used to model the level of stored energy in each UE's battery. Signal-to-interference ratio (SIR) coverage probability is evaluated.

PCP has been intensively investigated recently in the literature.
For instance, the authors in \cite{cluster_CSaha} considered networks in which the UE locations were modeled as a PCP with the BSs at the cluster centers. \cite{cluster_CSahaPCP} modeled a fraction of UEs and arbitrary
number of BS tiers alternatively with a PCP. In \cite{cluster_XJiang}  \cite{cluster_MAf} \cite{Cluster_CSaha_new}, the small-cell BSs were considered to be clustered and were modeled as PCPs. \cite{cluster_HTaba} provided a framework to analyze multi-cell uplink non-orthogonal multiple access (NOMA) systems where the UE locations form a PCP.
PCPs are also used in
device-to-device (D2D) networks, e.g.
\cite{cluster_MAf_Model}  \cite{Cluster_D2D_XWu} \cite{Cluster_D2D_MAfshang} \cite{Cluster_D2D_Esma} \cite{Milli_WYi}, where the locations of the D2D devices were modeled as PCPs.

PCP models have also been considered in the system-level analysis of UAV-assisted networks. In \cite{cluster_UAV_AHayajneh}, the UAVs were assumed to form a PCP with the destroyed macro BSs as the parent nodes. The downlink network performance, i.e. the SINR coverage probability, area spectral efficiency and energy efficiency, were investigated.  In \cite{UAV_Esma}, UAVs were considered as BSs serving the users. The UE locations were considered as PCPs. SINR coverage probability was investigated as the network performance metric. \cite{cluster_UAV_WYi} considered the UAV networks in millimeter wave (mmWave) communications. The UAVs were the parent nodes and were  3D deployed at same height, while the UEs were the daughter nodes and their locations  formed a Thomas cluster process. \cite{UAV_WYi} proposed a unified 3D spatial framework to evaluate the average performance of UAV-aided networks with mmWave communications. The UAVs and BSs were assumed to be PPP distributed and the UEs were distributed according to a PCP. During communication, a UAV received a message from a UE  in the uplink transmission and forwarded the message to a ground BS in the downlink transmission. The heights of the UAVs were all assumed to be the same.

%However, the aforementioned works neither modeled the UE locations as PCP, nor considered a joint downlink and uplink analysis scenario.

\subsection{Contributions}
In this paper, we consider UAV-assisted mmWave cellular networks, where the UEs are modeled according to a PCP and  downlink SWIPT scenario and uplink data transmission are jointly considered. The considered scenario can also address downlink energy transfer to low-power IoT devices and data collection from them using UAVs. The contributions of the paper are listed as follows:
\begin{itemize}
		\item A practical  UAV-assisted mmWave cellular network with PCP distributed UEs is addressed and studied in detail. In addition to ground BSs (GBSs), UAVs are also deployed according to a PPP distribution, and the UEs are considered to be clustered around the projections of UAVs according to PCPs. In this paper we specialize the PCP to Thomas cluster processes and Mat\'ern cluster processes. We characterize the complementary cumulative distribution function (CCDF) and the probability density function (PDF) of the distance from the typical UE to its own cluster center UAV, and other PPP-distributed UAVs and the GBSs. The CCDFs and PDFs are different from the existing studies on PCPs in two aspects: 1) the links being LOS or NLOS is taken into account; 2) the UAV height is incorporated.
    \item We jointly consider the downlink SWIPT scenario and uplink information transmission, where in downlink phase UEs both harvest energy and decode the information from the same received signal provided by the associated BS (either a UAV or a GBS), and in the uplink phase the UAVs collect data from their cluster member UEs. To the best of our knowledge,  this is one of the first studies that jointly consider the downlink SWIPT and uplink information transmission, i.e., the combination of SWIPT and WPCN, in UAV-assisted mmWave cellular networks. With this, we  provide a comprehensive analysis on this topic. For instance, the design of the time sharing parameter $\tau$ makes it possible to control the cooperation of the downlink and uplink phases depending on the mission of the UAVs.
	\item In the downlink phase, the largest received power association criterion is adopted and the power splitting technique is considered for the SWIPT scenario. Association probability and energy coverage of the proposed network are analyzed and general expressions are provided. Laplace transform of the interference is determined.  We also  define a realistic successful transmission probability to jointly address the energy coverage and SINR coverage performances of the considered network. The largest received power association criterion we used here is different from the prior work on UAV-assisted cellular networks with PCP models (e.g., \cite{cluster_UAV_AHayajneh} \cite{UAV_WYi}), since these studies adopted either the nearest association criterion or the random association criterion. And the largest received power association is more practical while being relatively more difficult to analyze. In addition, the adoption of the power splitting technique makes the model adaptive, since we can tune the power splitting component $\rho$ to control the trade-off between energy harvesting and information decoding. Even though the SINR coverage probability analysis is similar to the analysis in existing works that incorporate PCP models, the energy coverage analysis and the successful transmission probability analysis are the key novel components of our work and are substantially different from the SINR coverage probability. For instance, the successful transmission probability requires the characterization of the CCDF of the interference.
	\item In the uplink phase, each UAV is assumed to communicate with its cluster member UEs. According to the harvested energy of each UE in the downlink phase, UEs in the uplink phase are considered to be in either active mode or inactive mode. The Laplace transform of the inter-cell interference is again determined and the  SINR coverage probability is derived. In addition, the average uplink throughput subject to a constraint on the downlink throughput  is investigated to jointly address the downlink and uplink network performance. The Laplace transform of the inter-cell interference in the uplink analysis is non-trivial because of the PCP modeling. Moreover, due to the introduction of the of minimum harvested energy requirement and the consideration of the uplink average throughput optimization problem, the uplink phase interacts with the downlink phase, making the analysis more intricate.
	\item We provide an extension to multi-tier multi-height networks, demonstrating that our analysis is relatively broad and can be applied to more general networks. Additionally, we address the special case of noise-limited networks and derive closed-form expressions for the uplink SINR coverage probability and the optimal power splitting factor $\rho$, maximizing the downlink successful transmission probability.
	\item Via numerical results, several insightful characterizations are obtained. In particular, 1) it is shown that the system performance is improved when the cluster size becomes smaller; 2) optimal height of UAVs and optimal values of the power splitting parameter to maximize the system performance are determined; 3) in this network, the impact of the interference is negligible; 4) it is demonstrated that there exists an optimal time duration for the downlink phase that maximizes the average uplink throughput under a downlink throughput constraint; 5) the association criterion is shown to have impact on the SINR coverage performance; 6) it is observed that Thomas cluster processes and Mat\'ern cluster processes lead to similar network performance trends.
\end{itemize}

The rest of the paper is organized as follows: Network model and distance distributions are introduced in Section II and Section III, respectively. Section IV describes the UE association in both downlink and uplink phases. Section V focuses on the downlink coverage analysis, including the successful transmission probability. Section VI focuses on the uplink coverage analysis, including the average uplink throughput. We extend our analysis to a more general multi-tier multi-height UAV model and also investigate the special case of noise-limited networks in Section VII. In Section VIII, numerical and simulation results are presented to further investigate the network performance. Finally, a concluding summary is provided in Section IX. Proofs are relegated to the Appendix.

 \section{System Model}

In this section, we describe the considered UAV-assisted mmWave cellular network with PCP distributed UEs.

%\begin{figure}
%	\centering
%	\includegraphics[width=0.45 \textwidth]{UAV_assisted.eps} \\
%	\caption{\small A mmWave UAV-assisted cellular network. \normalsize}
%	\label{UAV_model}
%\end{figure}
\subsection{BS and UE deployment}
\subsubsection{UAV and GBS modeling}
The UAVs and GBSs are assumed to be distributed according to homogeneous PPPs $\Phi_U$ and $\Phi_G$ with densities $\lambda_U$  and $\lambda_G$, respectively. All UAVs and GBSs are assumed to be transmitting in a mmWave frequency band and have transmit powers $P_U$ and $P_G$, biasing factors to UEs $B_U$ and $B_G$, respectively. The biasing factor indicates the association preference of the tier, i.e. when we increase the $B$ of a tier, the UEs becomes more likely to be associated with the BS in that tier. All UAVs are assumed to be located at the same height $H$. We assume that all UAVs have enough energy resources to arrive at its 3D position in the air, communicate with UEs, and fly back.

\subsubsection{UE modeling}
\begin{figure}
	\begin{minipage}{0.45\textwidth}
		\centering
		\includegraphics[width=1\textwidth]{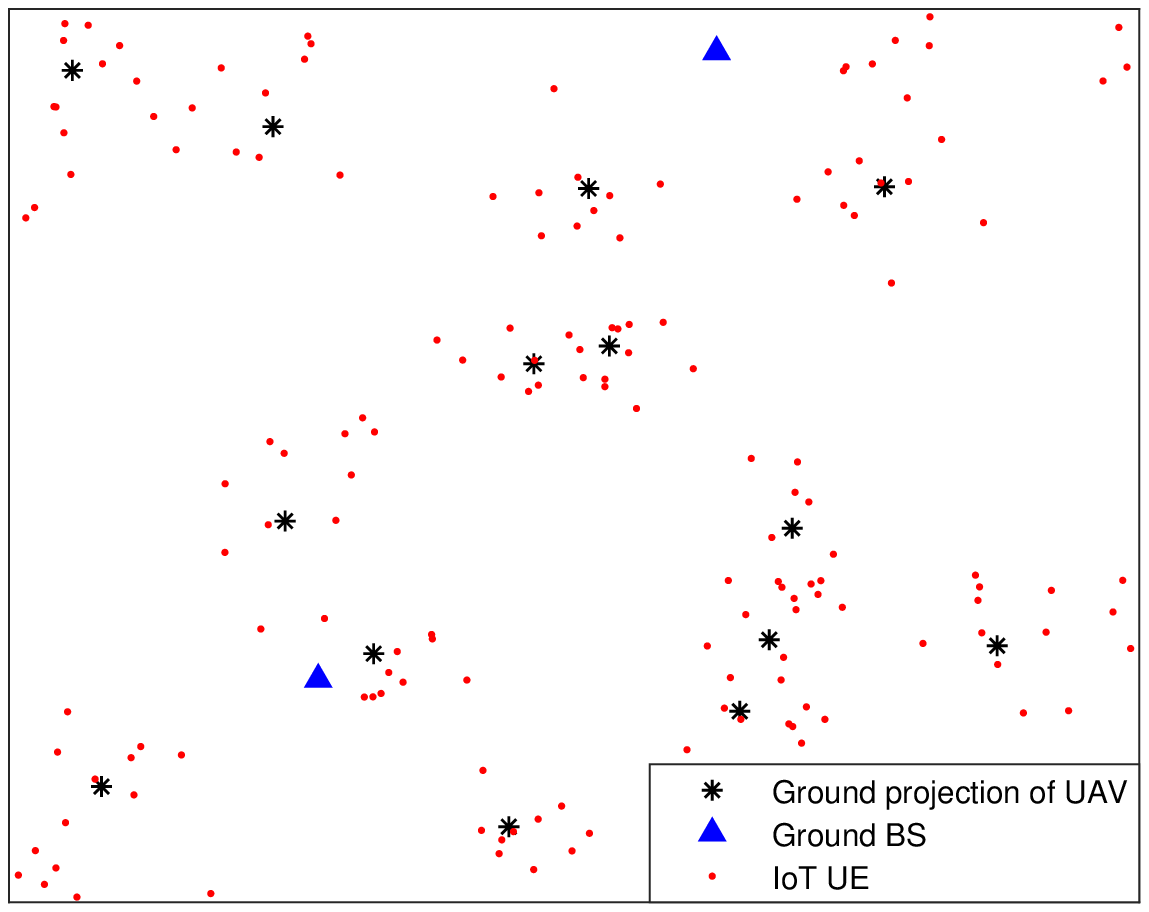} \\
		\subcaption{\scriptsize Thomas cluster process. }
	\end{minipage}
	\begin{minipage}{0.45\textwidth}
		\centering
		\includegraphics[width=1\textwidth]{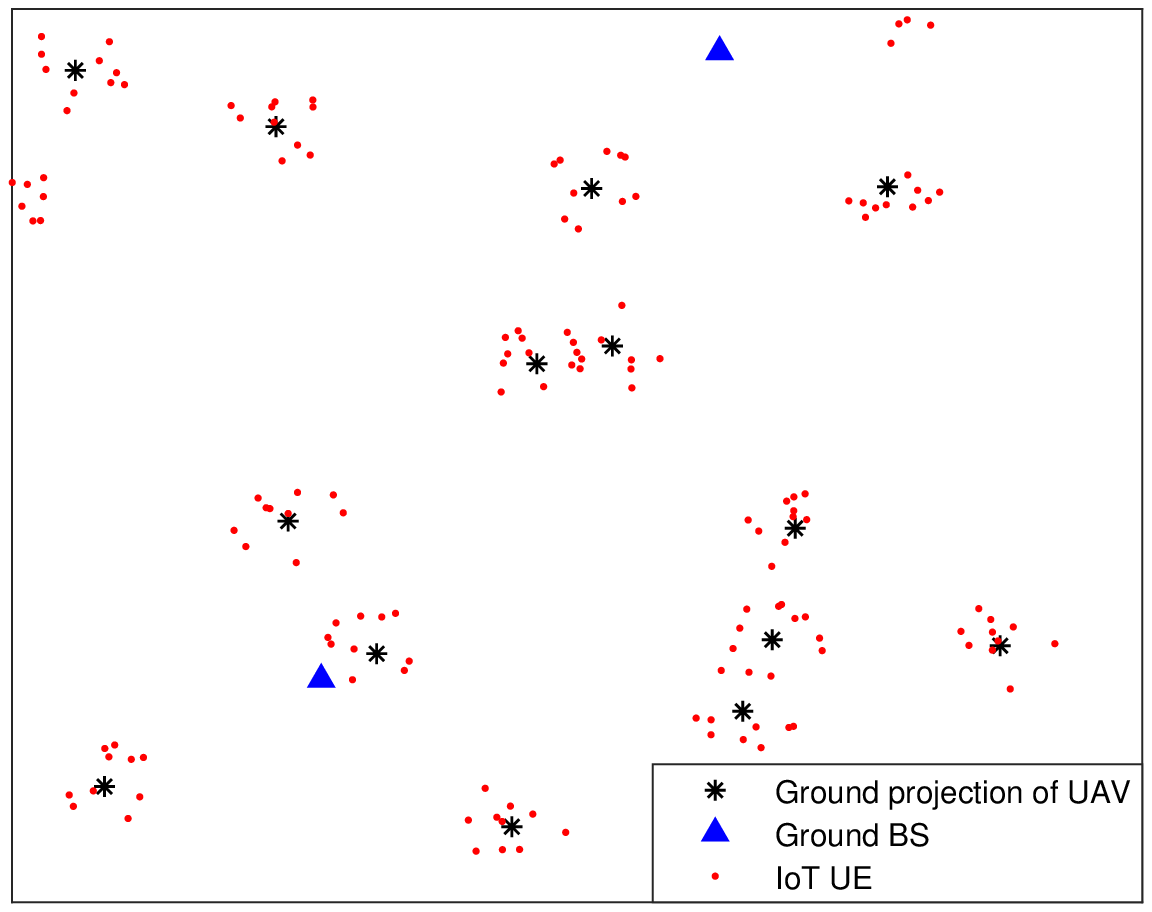}
		\subcaption{\scriptsize Mat\'ern cluster process.}
	\end{minipage}
	\caption{\small An illustration of the distributions of UAVs, GBSs and UEs.  \normalsize}
	\label{PCPmodel}
\end{figure}

The locations of the UEs are assumed to form a PCP denoted by $\Phi_u$, and the ground projections of the UAVs are the parent nodes. In this paper, we adopt two particular PCPs: (i)Thomas cluster processes, where UEs are symmetrically independently and identically distributed (i.i.d.) around the projections of the UAV locations on the ground according to a Gaussian distribution with variance $\sigma^2$; and (ii) Mat\'ern cluster processes, where UEs are distributed according to a uniform distribution within a circular disc of radius $R_c$. Sample realizations of spatially-distributed UEs along with UAVs and GBSs are depicted in Fig. \ref{PCPmodel}, considering both Thomas and Mat\'ern cluster processes.

Without loss of generality, in the downlink phase a random UE from a random cluster is chosen to be the typical UE and is assumed to be at the origin. To differentiate the distance from the typical UE to its cluster center UAV and the distance to other UAVs, we denote the cluster center as the $0^{th}$ tier UAV to the typical UE, and other UAVs and GBSs are the $1^{st}$ and $2^{nd}$ tier BSs, respectively. In the uplink phase, a UAV from a random cluster is chosen to be the typical BS. The descriptions of different tiers in the downlink phase are provided below in Table \ref{Table_tier}.

\begin{table}[htbp]
	\caption{Tiers in the Network}
	\centering
	\begin{tabular}{|l|p{3in}|}
		\hline
		\multicolumn{2}{|c|}{\textbf{Downlink phase}} \\ \hline
		$0^{th} $ tier  & The cluster center UAV of the typical UE \\ \hline
		$1^{st} $ tier  & Other PPP-distributed UAVs\\ \hline
		$2^{nd} $ tier  & The PPP-distributed GBSs \\ \hline
		$\Ksp =\{0,1,2\}$        &  The set of all tiers of UAVs and GBSs \\ \hline
	\end{tabular}
    \label{Table_tier}
\end{table}

%A mmWave UAV-assisted cellular network model is depicted in Fig. \ref{UAV_model}.

\subsection{Downlink and Uplink Transmission}
\begin{figure}
	\centering
	\includegraphics[width=0.5 \textwidth]{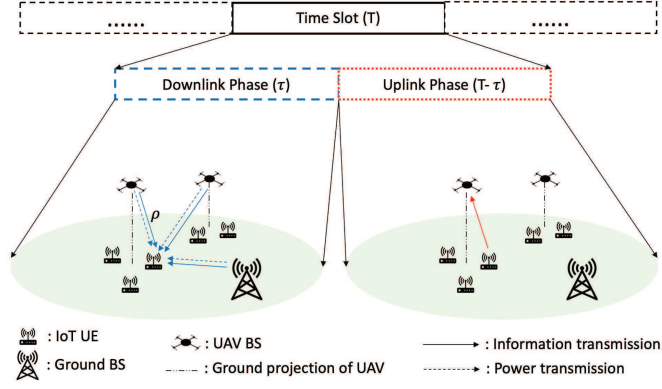} \\
	\caption{\small An illustration of the system model of a UAV-assisted mmWave  cellular network.\normalsize}
	\label{TransmissionModel}
\end{figure}

In this paper, we jointly consider downlink and uplink transmissions, where the UEs harvest energy and decode information from its downlink associated BS during downlink phase, and then send data to its cluster center UAV during uplink phase. The total time duration for downlink and uplink is assumed to be $T$ (seconds).  As shown in Fig. \ref{TransmissionModel}, each time frame of $T$ seconds is divided into downlink and uplink time slots with durations $\tau$ and $(T-\tau)$, respectively. In the downlink phase, SWIPT scenario is considered, and more specifically the power splitting technique is used. Employing this technique, the UEs can harvest energy and decode the information by splitting the received signal into two streams. The power splitting parameter that represents the power fraction used for information processing is denoted by $\rho$. It's assumed the UEs have enough battery storage to store the harvested energy. In the uplink phase, UEs use the harvested energy to send data to their cluster center UAVs. It is worth noting that when $\tau=T$, our model specializes to a downlink SWIPT network. Additionally, when $\rho=0$, we recover the network model with downlink energy harvesting and uplink data transmission (i.e. the WPCN scenario).

\subsection{Channel Modeling}
Link between a UE and a BS can be either LOS or non-LOS (NLOS). The path loss model is formulated as
\begin{align}
L^s_j(r)=
\begin{cases}
\kappa^L_j r^{\alpha^L_j}(r) & \text{with prob. } \quad p^L_j(r)\\
\kappa^N_j r^{\alpha^N_j}(r) &  \text{with prob. }  \quad p^N_j(r)=(1-p^L_j(r))
\end{cases}\label{eq:pathlossformulations}
\end{align}
where $j\in \Ksp$, $s\in\{\los,\nlos\}$ and superscripts $L$ and $N$ indicate LOS and NLOS, respectively.  In the $j^{th}$ tier, $\alpha^L_j, \alpha^N_j$ are the path loss exponents for LOS and NLOS links, respectively, $\kappa^L_j , \kappa^N_j$ are intercepts of the LOS and NLOS path loss formulas, respectively, and $p^L_j(r)$ is the probability that the link has a LOS transmission at distance $r$.
\subsubsection{Air to Ground}
Similarly as in \cite{UAV_AHourani}, we formulate the probability of the LOS link between the UAVs and the UEs as
\begin{align}
			p^L_U(r)=\frac{1}{1+C \exp(-B(\theta-C))}
\end{align}
where $\theta=\frac{180}{\pi} \arcsin (\frac{H}{r})$ is the elevation angle, $H$ denotes the height of the UAVs, and $B$ and $C$ are specific constants that depend on the environment (rural, urban, dense urban, etc.). Note that since the $0^{th}$ and $1^{st}$ tier BSs are UAVs, we have $p^L_0(r) = p^L_1(r) = p^L_U(r)$ in the path loss formulations in (\ref{eq:pathlossformulations}).

\subsubsection{Ground to Ground}
%Different propagation characteristics like LOS and NLOS are taken into account.
Considering mmWave transmissions, similar to \cite{Milli_CGaliotto} and \cite{Milli_Bai}, we formulate the probability of LOS link between the GBSs and the UEs as
\begin{align}
				 p^L_2(r) = p^L_G(r)=e^{-\epsilon r}
\end{align}
where  $\epsilon$ is a constant that depends on the geometry and density of the building blockage process. The smaller $\epsilon$ is, the sparser the environment will be.

It's worth noting that since we distinguish the links between the UEs to the GBS as either LOS or NLOS,  we assume the BSs in the $1^{st}$ and $2^{nd}$ tiers are divided into two independent PPPs $\Phi_j^s$
for $s\in\{\los,\nlos\}$.%with density $E_r[\lambda_j p^s_j(r)]$.

\subsection{Antenna gain}
Considering that directional transmissions are performed in mmWave communications, we address a sectored antenna model in this paper, where $M_*$ and $m_*$ are the the main lobe gain and side lobe gain, respectively, and $*\in \{b,u \}$ denotes the BS side or the UE side. We further assume that the antenna gain between the UE and the serving BS can achieve the maximum antenna gain $G_0=M_b M_u$. On the other hand, we assume the beam direction of the interfering links is modeled as uniformly distribution over $[0, 2\pi)$.  Therefore, we can formulate the antenna gain of an interfering link as \cite{Milli_Bai}
\begin{align}
G=
\begin{cases}
M_b M_u & w.p. \quad p_{M_b M_u}=(\frac{\theta_b}{2\pi})(\frac{\theta_u}{2\pi})\\
M_b m_u & w.p. \quad p_{M_b m_u}=(\frac{\theta_b}{2\pi})(1-\frac{\theta_u}{2\pi})\\
m_b M_u & w.p. \quad p_{m_b M_u}=(1-\frac{\theta_b}{2\pi})(\frac{\theta_u}{2\pi})\\
m_b m_u & w.p. \quad p_{m_b m_u}=(1-\frac{\theta_b}{2\pi})(1-\frac{\theta_u}{2\pi}),
\end{cases}
\end{align}
where $\theta_*$ for $*\in \{b,u \}$ denotes the main lobe beamwidth.

\subsection{Small-scale Fading}
Nakagami-$m$ fading is a general fading model suitable under various conditions \cite{UAV_LZhou},  and hence we assume all transmission links experience independent Nakagami-$m$ fading\footnote{Note that Nakagami fading specializes to Rayleigh fading when $m=1$.}.
Denoted by $h_s$, the small-scale fading gains (i.e., magnitude-squares of fading coefficients) follow Gamma distributions $h_l \sim \Gamma(N_l, 1/N_l)$ for LOS, while $h_n \sim \Gamma(N_n, \frac{1}{N_n})$ for NLOS, where $N_l, N_n$ are the Nakagami fading parameters for LOS and NLOS links, respectively, and are assumed to be positive integers.

A summary of notations is provided in Table \ref{Table_notation}. The abbreviations of symbols that are used to simplify the expressions are not included.
\begin{table}[htbp]
	\caption{Table of Notations}
	\centering
	\begin{tabular}{|l|p{4.5in}|}
		\hline
		\footnotesize \textbf{Notations} &  \footnotesize  \textbf{Description}  \\ \hline
		
		\scriptsize$\Phi_U,\lambda_U,\Phi_G,\lambda_G$  &  \scriptsize  PPP of the UAVs (named as the $1^{st}$ tier BS), the density of $\Phi_U$, PPP of the GBSs (named as the $2^{nd}$ tier BS), the density of $\Phi_G$.\\\hline
		\scriptsize$\Phi_u$ &  \scriptsize  PCP of the UE locations.\\\hline
		\scriptsize$\sigma, R_c$  &  \scriptsize  Cluster size, denoted by the standard deviation, if $\Phi_u$ is Thomas cluster process, and the radius of the cluster, if $\Phi_u$ is Mat\'ern cluster process. \\\hline				
		\scriptsize$P_U,B_U,P_G,B_G$  &\scriptsize  Transmit power and biasing factor of the UAVs and the GBSs. \\ \hline
		\scriptsize$P_t^{UL}$  &\scriptsize  The transmit power of the UEs. \\ \hline
		\scriptsize$H$  &\scriptsize  The height of the UAVs. \\ \hline
		\scriptsize$T,\tau,\rho$  &\scriptsize  The total time duration, the time duration for downlink phase, the power fraction used for information processing in downlink phase. \\ \hline
		\scriptsize$L^{s}_j,p^s_j$  &\scriptsize  Path loss and the probability of $s\in\{\los,\nlos\}$ transmission in the $j\in \Ksp $-th tier.  \\ \hline
		\scriptsize$\alpha^s_j, \kappa^s_j$  &\scriptsize  The path loss components, and the path loss intercepts. \\ \hline
		\scriptsize$\theta, \epsilon$  &\scriptsize  The elevation angle of the UAV, and a constant that depends on the geometry for the ground to ground LOS transmission. \\ \hline
		\scriptsize$M_*,m_*, \theta_*$  &\scriptsize  Main lobe gain, side lobe gain, and the beamwidth of the main lobe where $*$ is BS or UE.\\ \hline
		\scriptsize$G, p_G$  &\scriptsize Antenna gain and the corresponding probability. \\ \hline
		\scriptsize$G_0$  &\scriptsize MM, which is the antenna gain of the main link. \\ \hline
		\scriptsize$h, N_s$   &\scriptsize Small-scale fading gain, the fading parameters for LOS/NLOS. \\ \hline
		\scriptsize$\sigma^2_{n}, \sigma^2_{c}$   &\scriptsize The thermal noise variance and the noise factor variance due to the conversion of the received bandpass signal to baseband. \\ \hline
		\scriptsize$R_0,R_U,R_G$  &\scriptsize  The distance from a UE to its cluster center UAV, the nearest $1^{st}$ tier UAV and the nearest GBS. \\ \hline
		\scriptsize$R_{UU}$  &\scriptsize  The distance from a UE to other cluster center UAV. \\ \hline
		\scriptsize$D^s_0,D^s_U,D^s_G$  &\scriptsize  The probabilities that the typical UE has a LOS/NLOS $0^{th}$ tier UAV, at least one LOS/NLOS UAV, or at least one LOS/NLOS GBS around. \\ \hline
		\scriptsize$P_m, I$  &\scriptsize The received power of the main link, and the interference. \\ \hline
		\scriptsize$\cL_I (a)$  &\scriptsize The Laplace transform of $I$ at evaluated at $a$. \\ \hline
		\scriptsize$E^{hv}, \sinr$  &\scriptsize The harvested energy and the signal-to-interference-pluse-noise ratio. \\ \hline
		\scriptsize $q(0,\cdot)$   &   \scriptsize Exclusion disc, inside which no interference exists in the dwonlink phase.  \\ \hline
		\scriptsize $\gamma_E, \gamma_{sinr}$, $\gamma^{UL}$ &  \scriptsize The energy coverage probability threshold, the SINR coverage probability threshold in the dwonlink phase, and the SINR coverage probability threshold in the uplink phase.\\ \hline
		\scriptsize  $p_{active}$ &   \scriptsize The probability that the UE is active in the uplink phase. \\ \hline		
		\scriptsize $A_j , A_{j,s}$   &   \scriptsize Association probability with a BS or a LOS/NLOS BS in the $j^{th}$ tier in the downlink phase. \\ \hline
		\scriptsize $P_E , P_{SINR}, P_{ST}$   &   \scriptsize The energy coverage probability, the SINR coverage probability and the successful transmission probability in the dwonlink phase. \\ \hline
		\scriptsize $P^{UL}_{SINR}$, $R^{UL}$ &   \scriptsize The SINR coverage probability and the average throughput in the uplink phase. \\ \hline				
	\end{tabular}
	\label{Table_notation}
\end{table}\normalsize

\section{Distance Distributions}
\begin{figure}
	\centering
	\includegraphics[width=0.45 \textwidth]{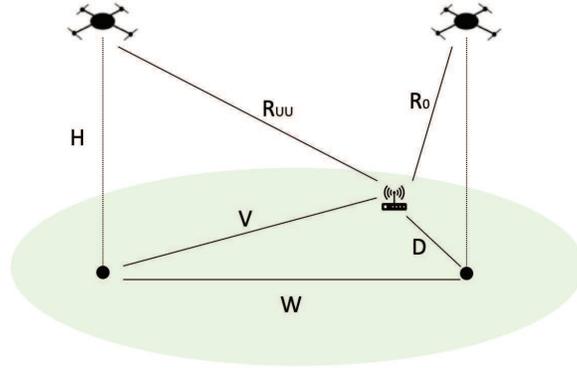} \\
	\caption{\small An illustration of difference distance in the network.\normalsize}
	\label{DistanceModel}
\end{figure}

In this section, we characterize the CCDF and the PDF of the distance from the typical UE to UAVs and GBSs in each tier. Fig. \ref{DistanceModel} provides an illustration of different distances. These distance distributions are subsequently employed to characterize the association probability in Section IV and the networks performance metrics in Section V and VI.

\subsection{The distance $R_0$ from the typical UE to the $0^{th}$ tier UAV}
The distance from the typical UE to the projection of its cluster center UAV on the ground is denoted as $D$. Then the distribution of $D$ can be expressed for different PCPs  as follows \cite{cluster_MAf_Model}:
\subsubsection{Thomas cluster process}
\begin{align}
&\text{CCDF: }\qquad \overline{F}_{D}(x)=\exp \left( \frac{-{x^2}}{2{\sigma^2}}\right)    , \label{CCDF_Th} \\
&\text{PDF:    }\qquad f_{D}(x)=\frac{x}{{\sigma^2}}\exp\left(\frac{-{x^2}}{2{\sigma^2}}\right)   , \label{PDF_Th}
\end{align}
where $x\geq 0$ and $\sigma^2$ is the variance of the UE distribution.
\subsubsection{Mat\'ern cluster process}
\begin{align}
	&\text{CCDF: } \hspace{0.22in} \overline{F}_{D}(x)= \left(1- \frac{x^2}{R_c^2}\right) \mathbb{1}(0 \leq x \leq R_c),\\
	&\text{PDF:    }\qquad f_{D}(x)= \frac{2x}{R_c^2}\mathbb{1}(0 \leq x \leq R_c),
\end{align}
where $0 \leq x \leq R_c$, $R_c$ is the radius of the cluster and $\mathbb{1}(\cdot)$ is the indicator function.

\begin{Lem}
	Given that the link between the typical UE and its cluster center UAV is in $s\in\{\los,\nlos\}$ transmission, the CCDF and PDF of $R^s_0$ can be expressed as follows: \\
    (i)	Thomas cluster process:
	\begin{align}
	&\text{CCDF: }  \qquad
	\Fbar_{R^s_{0}}(x) = \int_{\sqrt{x^2-H^2}}^{\infty} p^s_U(\sqrt{d^2+H^2}) f_D(d) \, \dsp d / D^s_0,  \qquad (x\geq H) \\
	& \hspace{0in } \text{PDF: }  \qquad
	 f_{R^s_{0}} (x) =\frac{x p^s_U(x)}{{\sigma^2 D^s_0}}\exp\left(\frac{H^2-x^2}{2{\sigma^2}}\right) \qquad (x\geq H) ,
%    f_{R^s_{0}} (x) = \frac{x p^s_U(x)}{ D^s_0\sqrt{x^2-H^2}} f_D(\sqrt{x^2-H^2})
	\end{align}
	(ii) Mat\'ern cluster process:
	\begin{align}
	&\text{CCDF: } \hspace{0.22in}
	\Fbar_{R^s_{0}}(x) = \int_{\sqrt{x^2-H^2}}^{R_c} p^s_U(\sqrt{d^2+H^2}) f_D(d) \, \dsp d / D^s_0, \qquad (H \leq x \leq \sqrt{H^2+R_c^2} ),\\
	& \hspace{0in } \text{PDF: }  \qquad
	f_{R^s_{0}} (x) =\frac{2 x p^s_U(x)}{{R_c^2 D^s_0}}, \hspace{2.22in}(H \leq x \leq \sqrt{H^2+R_c^2} ),
	\end{align}
	where $D^s_0=\int_{0}^{\infty} p^s_U(\sqrt{d^2+H^2}) f_{D} (d)\dsp d$ is the probability that the link is in $s \in \{ \los, \nlos \}$ transmission.

\end{Lem}

\emph{Proof:} See Appendix \ref{Proof_0TH_CCDF}.

Therefore, we can obtain the CCDF and PDF of $R_{0}$ as follows:
\begin{align}
 & \Fbar_{R_{0}} (x) = \sum_s D^s_0 \Fbar_{R^s_{0}} (x) %=\int_{\sqrt{x^2-H^2}}^{\infty} f_D(d) dd = \Fbar_D({\sqrt{x^2-H^2}}) \notag \\
=\begin{cases}
 			\exp \left( \frac{H^2-x^2}{2{\sigma^2}}\right )  \qquad (x\geq H) \hspace{1.2in} \text{for Thomas cluster process}\\
 			 1- \frac{x^2-H^2}{R_c^2} \qquad (H \leq x \leq \sqrt{H^2+R_c^2} )\qquad \text{for Mat\'ern cluster process}
\end{cases}\\
&	f_{R_{0}} (x) = - \frac{d \Fbar_{R_{0}} (x)  }{dx}
=\begin{cases}
		\frac{x}{{\sigma^2}}\exp\left(\frac{H^2-x^2}{2{\sigma^2}}\right) \qquad (x\geq H) \qquad \text{for Thomas cluster process} \\
		\frac{2x}{R_c^2}  \qquad (H \leq x \leq \sqrt{H^2+R_c^2} )\qquad \text{for Mat\'ern cluster process}.
\end{cases}
\end{align}

%If Mat\'ern cluster process is considered, the UEs are clustered around the base stations according to a uniform distribution. Then the CCDFs and PDFs  of $D$, $R_0$ and $R_0^s$ can be expressed as
%\begin{align}
%	&\text{CCDF: } \hspace{0.22in} \overline{F}_{D}(x)= 1- \frac{x^2}{R_c^2}, \hspace{2.32in} (0 \leq x \leq R_c ), \\
%	&\text{PDF:    }\qquad f_{D}(x)= \frac{2x}{R_c^2}, \hspace{2.66in} (0 \leq x \leq R_c ),  \\
%	&\text{CCDF: } \hspace{0.22in}\overline{F}_{R_0}(x)= 1- \frac{x^2-H^2}{R_c^2}, \hspace{1.89in} (H \leq x \leq \sqrt{H^2+R_c^2} ),\\
%	&\text{PDF:    }\qquad f_{D}(x)= \frac{2x}{R_c^2},  \hspace{2.66in} (H \leq x \leq \sqrt{H^2+R_c^2} ),\\
%	&\text{CCDF: } \hspace{0.22in}
%	\Fbar_{R^s_{0}}(x) = \int_{\sqrt{x^2-H^2}}^{R_c} p^s_U(\sqrt{d^2+H^2}) f_D(d) dd / D^s_0, \qquad (H \leq x \leq \sqrt{H^2+R_c^2} ),\\
%	& \hspace{0in } \text{PDF: }  \qquad
%	f_{R^s_{0}} (x) =\frac{2 x p^s_U(x)}{{R_c^2 D^s_0}}, \hspace{2.22in}(H \leq x \leq \sqrt{H^2+R_c^2} ),
%\end{align}
%where $R_c$ is the radius of the clusters.

\subsection{The distance $R^s_U$ from the typical UE to the nearest LOS/NLOS UAV from the $1^{st}$ tier}
\begin{Lem}
				Given that the typical UE can observe at least one LOS/NLOS UAV in the $1^{st}$ tier, the CCDF and PDF of $R^s_U$ can be expressed as follows:
				\begin{align}
				&\text{CCDF: }\quad           \Fbar_{R^s_U}(x) = e^{-2 \pi \lambda_U \int_H^x t p^s_U(t) dt } /{D^s_U},  \\
				&\text{PDF:    }\quad           f_{R^s_U}(x) = 2 \pi \lambda_U x p^s_U(x)  e^{-2 \pi \lambda_U \int_H^x t p^s_U(t) dt } /{D^s_U},
				\end{align}
				where $x\geq H$, $D^s_U=1-e^{-2 \pi \lambda_U \int_H^\infty t p^s_U(t) dt}$ is the probability that the typical UE has at least one LOS/NLOS UAV around.

 \end{Lem}

\emph{Proof:} See Appendix \ref{Proof_UAV_CCDF}.

\subsection{The distance $R^s_G$ from the typical UE to the nearest LOS/NLOS GBS from the $2^{nd}$ tier}
Given that the typical UE can observe at least one LOS/NLOS BS in the $2^{st}$ tier, the CCDF and PDF of $R^s_G$ can be determined from \cite[Lemma 1]{Milli_Bai} as follows:
\begin{align}
&\text{CCDF: }\qquad           \Fbar_{R^s_G}(x) =  e^{-2 \pi \lambda_G \int_0^x t p^s_G(t) dt } /{D^s_G},  \\
&\text{PDF:    }\qquad           f_{R^s_G}(x) = 2 \pi \lambda_G  x p^s_G(x) e^{-2 \pi \lambda_G \int_0^x t p^s_G(t) dt } /{D^s_G},
%          f_n(x) &= 2 \pi \lambda B^{-1}_n x (1-p_l(x)) e^{-2 \pi \lambda \int_0^x t (1-p_l(t)) dt } \quad x \geq 0 \\
%          \Fbar_n(x) &= B^{-1}_n e^{-2 \pi \lambda \int_0^x t (1-p_l(t)) dt } \quad x \geq 0
\end{align}
where $x\geq 0$, $D^s_G=1-e^{-2 \pi \lambda_G \int_0^\infty t p^s_G(t) dt}$ is the probability that the typical UE has at least one LOS/NLOS GBS around.

\subsection{The distance $R_{UU}$ from a UE to the other cluster center UAV}
 For Thomas cluster process, the PDF of the distance $V$ from a UE to the ground projection of other cluster center UAV, given the distance $W$ from the UE's cluster center UAV to the corresponding UAV, can be expressed as \cite{cluster_MAf_Model}
\begin{align}
			f_{V} (v|w)=\frac{v}{\sigma^2} \exp\left(- \frac{v^2+w^2}{2 \sigma^2}\right) I_0\left(\frac{vw}{\sigma^2}\right),
\end{align}
where $I_0(\cdot)$ is the modified Bessel function with order zero.
For Mat\'ern cluster process, the PDF can be expressed as  \cite{PCP_JTang}
\begin{align}
	f_{V} (v|w)= \frac{2v}{\pi R_c}\arccos\frac{v^2+w^2-R^2_c}{2vw} \indictor\left(|R_c-w|\leq v \leq R_c+w \right) + \frac{2v}{R^2_c}\indictor(v<R_c-w).
\end{align}
Then the PDF of $R_{UU}$ can be obtained as
\begin{align}
	 f_{R_{UU}}(x|w) = \frac{x}{\sqrt{x^2-H^2}} f_{V} (\sqrt{x^2-H^2}|w).
\end{align}

\section{User Association}
In this section, we focus on the downlink and uplink UE association criterion, and also provide the downlink association probability of each tier, from which we can determine how the UEs connect with the UAVs and GBSs.
%In this section, we first investigate the association probability of each tier, then analyze the energy and SINR coverage of the proposed system and finally provide general expressions.

\subsection{Downlink association}
In the downlink phase, UEs need to harvest energy and decode the information from the associated BS (e.g., a UAV or a GBS). The strongest biased average power association criterion \cite{Hetero_SSingh}\cite{Milli_XWang_journal} is utilized, i.e. the UEs are assumed to be associated with the BS providing the strongest long-term biased average received power.
 Since the antenna gain of the main link is assumed to achieve the maximum value $G_0$, the received power of the main link can be expressed as
\begin{align}
          P_m&= \argmax_{j \in \Ksp, i\in \Phi} P_j G_0 B_j L^{-1}_{ji}
          \overset{(a)}{=} \argmin\limits_{j \in \Ksp, s } P_j G_0 B_j (\kappa^s_j (r^s_j)^{\alpha^s_j})
          %=\argmin_{j \in \{0,1,2\}, i\in \Phi} \kappa_{}
\end{align}
where $r^s_j$ is the distance from the typical UE to the nearest LOS/NLOS BS in the $j^{th}$ tier, and (a) follows from the fact that in each tier the transmit power and the biasing factor are the same, and therefore the maximum received power is from the nearest  LOS/NLOS BS.

\begin{Lem}
The probability that the typical UE is associated with a LOS/NLOS BS in the $j^{th}$ tier is given by
	\begin{align} \label{AP}
					A_{j,s}=
					\begin{cases}
					 \E_{R^s_{0}} \left[  D_0^s \prod\limits_{k} \prod\limits_{b} D_k^b  \Fbar_{R_k^b} \left(Q_{k0}^{sb} (r_0)\right)\right],  \hspace{2.8in}\text{for }j=0, \\
					 D_j^s\E_{R^s_{j}}  \Bigg[D_j^{s'} \Fbar_{R^{s'}_{j}}\left(Q_{jj}^{ss'} (r_j)\right)
					 \hspace{0in} \left(\sum\limits_b D_0^b \Fbar_{R^b_{0}}\left(Q_{0j}^{sb}(r_j) \right)\right) \prod\limits_b D_k^b  \Fbar_{R_k^b} \left(Q_{kj}^{sb} (r_j)\right) \Bigg],  \hspace{0.10in} \text{for }j=1,2,
					    %    \hspace{2.10in} \text{for }j=1,2,
					\end{cases}
	\end{align}
	where $s,s',b \in \{\los, \nlos \}$, $s'\neq s$, $k=1,2$, $Q_{kj}^{sb}(r)=\left(\frac{P_k B_k \kappa_j^s}{P_j B_j \kappa_k^b} r^{\alpha^s_j}\right)^{\frac{1}{\alpha^b_k}}$, $D^s_j$ and $\Fbar_{R^s_j}(x)$ are given in Section III.

\end{Lem}

	\emph{Proof:} See Appendix \ref{Proof_AP}.

\begin{Rem}
	In order to characterize the link level performance of the UAV-assisted network, we will need to find the distance distribution give that link. Therefore, given that the typical UE is associated with a LOS/NLOS BS in the $j^{th}$ tier, the PDFs of the distances from the typical UE to the associated BS can be expressed as follows:
	\begin{align}
	\fhat_{R^s_{j}}(x) =
	\begin{cases}
	\frac{f_{R^s_0}(x)}{A_{0,s}} D_0^s \prod\limits_{k} \prod\limits_{b} D_k^b  \Fbar_{R_k^b} \left(Q_{k0}^{sb} (x)\right) , \hspace{2.72in}\text{for }j=0,\\
	\frac{f_{R^s_j}(x)}{A_{j,s}} D_j^{s} D_j^{s'} \Fbar_{R^{s'}_{j}}\left(Q_{jj}^{ss'} (x)\right)
	\hspace{0in} \left(\sum\limits_b D_0^b \Fbar_{R^b_{0}}\left(Q_{0j}^{sb}(x) \right)\right) \prod\limits_b D_k^b  \Fbar_{R_k^b} \left(Q_{kj}^{sb} (x)\right),
	\hspace{0.1in} \text{for }j=1,2.
	\end{cases}
	\label{PDF_cond}
	\end{align}
	And the proof follows the same way as the association probability in Appendix \ref{Proof_AP}.
\end{Rem}

\subsection{Uplink association}
In the uplink phase, each UAV aims to collect data from one cluster, and hence the UAV is assumed to communicate with its own cluster member UEs.  It is further assumed that different UEs in one cell are served using orthogonal resources, and hence no intra-cell interference exists. UEs from other clusters can inflict interference. It is worth noting that UEs may not be associated with the same BSs in downlink and uplink phases, due to the adoption of the strongest biased average power association criterion in the downlink phase.

\section{Downlink Coverage analysis }
In this section, we first investigate the interference in the downlink phase, then analyze the network performance by the energy coverage and SINR coverage of each tier. Finally, we provide a successful transmission probability which can jointly consider both energy coverage and SINR coverage and can represent the downlink performance of the UAV-assisted cellular network.

\subsection{Interference}
Since the typical UE is assumed to be served by a BS which provides the largest biased received power $P_m$, then if a UE is associated with a LOS/NLOS BS from the $j^{th}$ tier at distance $r$, there exists an exclusive disc $q (0,Q_{kj}^{sb} (r))$ in which no interfering BS exists. Therefore, the experienced interference at the typical UE can be expressed as follows:
\begin{align}
&I = I_0+I_1+I_2   \\
&I_0=P_0 G h_0 (\kappa^b_U r_0^{\alpha^b_U} )^{-1} \\
&I_k=  \sum\limits_{b}\sum\limits_{i\in \Phi^b_k\backslash q} P_k G_i h_k (\kappa^b_k r_{k,i}^{\alpha^b_k})^{-1}
\end{align}
where $b\in\{\los,\nlos\}$, $k=1,2$, $r_0$ denotes the distance from the UE to its cluster center, and $r_{k,i}$ stands for the distance from the UE to the $i^{th}$ BS in the $k^{th}$ tier. It is worth noting that when the serving BS is from the $0^{th}$ tier, we have $I_0 =0$, since there's only one BS in this tier.

\subsection{Harvested Energy and Signal to Interference Plus Noise Ratio (SINR) }
Since power splitting technique is employed with parameter $\rho$, the total harvested energy of the typical UE in the downlink phase can be expressed as
\begin{align}
E^{hv}_{j,s}=\tau(1-\rho)(P_m + I)
\end{align}
where $\tau$ is the time duration used for downlink phase, and $P_{m}=P_j G_0 h_j (\kappa^s_j r^{\alpha^s_j})^{-1} $ denotes the received power of the main link from the serving BS. We neglect the additive white Gaussian noise (AWGN) term in energy harvesting. It is worth noting that we assume linear energy harvesting, and the case of non-linear energy harvesting remains as future work.

Moreover, the experienced SINR at the typical UE can be expressed as
\begin{align}
\sinr_{j,s} &=\frac{\rho P_{m}}{\sigma_c^2 + \rho(\sigma^2_n+I)} = \frac{P_{m}}{ \frac{\sigma_c^2}{\rho} + \sigma^2_n+I}
\end{align}
where $\sigma^2_n$ is the variance of the Gaussian thermal noise component and $\sigma^2_c$ is the noise factor due to the conversion of the received bandpass signal to baseband.

\subsection{Energy Coverage Probability}
The energy coverage probability can be defined as the probability that the harvested energy is larger than a certain threshold $\gamma_E>0$. Therefore, given the event $S_{j,s} =$ \{The typical UE is associated with a LOS/NLOS BS in the $j^{th}$ tier\}, the conditional energy coverage can be expressed as
\begin{align}
		P^c_{E_{j,s}}(\rho,\tau, \gamma_E)=\prob(E^{hv}_{j,s}>\gamma_E| S_{j,s}).
\end{align}
Hence, the energy coverage probability of the entire network can be obtained by
\begin{align}
		& P_E(\rho,\tau, \gamma_E)=\sum\limits_{j\in\Ksp} P_{E_{j}}(\rho,\tau,\gamma_E) = \sum\limits_{j\in\Ksp}\sum\limits_{s} P^c_{E_{j,s}}(\rho,\tau,\gamma_E) A_{j,s}
\end{align}
where $A_{j,s}$ is the association probability given in (\ref{AP}), and $P_{E_{j}}(\rho,\tau,\gamma_E)=\sum_s P^c_{E_{j,s}}(\rho,\tau,\gamma_E) A_{j,s}$ is the energy coverage of tier $j$.

\begin{Theo}
	Conditioned on that the typical UE is associated with a LOS/NLOS BS in the $j^{th}$ tier, the energy coverage probability can be expressed as follows:
	\begin{align} \label{EC}
			P^c_{E_{j,s}}(\rho,\tau,\gamma_E)=
			\begin{cases}
%					\sum\limits_{n=0}^N (-1)^n \Big(\substack{N \\ \\n}\Big) \int\limits_{0}^{\infty}\zeta^{G_0}_{j,s}(r) \fhat_{R^s_0}(r) e^{-\sum\limits_k \sum\limits_b \sum\limits_G \chi^{sbG}_{k0}(r,\zeta)} dr
					\sum\limits_{n=0}^N (-1)^n \Big(\substack{N \\ \\n}\Big) \int\limits_{H}^{\infty}\zeta^{G_0}_{j,s}(r) \fhat_{R^s_0}(r) \prod\limits_k \cL_{I_k}(\ahat) dr
					\hspace{1.55in} \text{for } j=0,\\
%					\sum\limits_{n=0}^N (-1)^n \Big(\substack{N \\ \\n}\Big)
%					\int\limits_{B_{d1}}^{\infty}\zeta^{G_0}_{j,s}(r) \fhat_{R^s_j}(r) \left(\sum\limits_b \sum\limits_G \delta^{sbG}_j(r,\zeta) \right)
%					\hspace{0in} e^{ - \sum\limits_k \sum\limits_b \sum\limits_G \chi^{sbG}_{kj}(r,\zeta)} dr \qquad \text{for } j=1,2,
					\sum\limits_{n=0}^N (-1)^n \Big(\substack{N \\ \\n}\Big) \int\limits_{B_{d1}}^{\infty}\zeta^{G_0}_{j,s}(r) \fhat_{R^s_j}(r) \cL_{I_0}(\ahat) \prod\limits_k \cL_{I_k}(\ahat)dr \hspace{1in} \text{for } j=1,2,
			\end{cases}
	\end{align}
	where $s,b\in \{\los,\nlos\} $, $\ahat=\frac{a n \tau (1-\rho)}{\gamma_E}$, $a=N(N!)^{-\frac{1}{N}}$, $B_{d1}=H$ for $j=1$, $B_{d1}=0$ for $j=2$, $\label{zeta} \zeta^G_{j,s}(r)=\left(1+\ahat P_j G (\kappa^s_j r^{\alpha^s_j} N_s)^{-1} \right)^{-N_s}$, and $\fhat_{R^s_j}(r)$ is the conditional PDF of distances given in (\ref{PDF_cond}). The Laplace transforms of the interference can be expressed as follows:
 	  \begin{align}
%			&\label{zeta} \zeta^G_{j,s}(r)=\left(1+\ahat P_j G (\kappa^s_j r^{\alpha^s_j} N_s)^{-1} \right)^{-N_s} \\
%			&\label{chi} \chi^{sbG}_{kj}(r,\zeta)=2 \pi \lambda_k\int\limits_{B_{d2}}^{\infty} \left(1-\zeta^G_{k,b}(x)\right)p_G p^b_k(x) x dx\\
%			&\label{delta} \delta^{sbG}_{j}(r,\zeta)= \int\limits_{\max(H,Q^{sb}_{0j}(r))}^{\infty} p_G \zeta^G_{0,b}(x) \frac{f_{R^b_0}(x)}{F_{R^b_0}(Q^{sb}_{0j}(r))} d_x
			&\cL_{I_0}(\ahat)=\sum\limits_{G} \sum\limits_{b}  \int_{\max\{H,  Q_{0j}^{sb}(r)\}}^\infty \frac{p_G f_{R^b_{0}}(r_0) dr_0}{\left(1+\ahat P_0 G (\kappa^b_0 r_0^{\alpha^b_0} N_b)^{-1}\right)^{N_b} \Fbar_{R^b_{0}}(Q_{0j}^{sb}(r))}  \\
			&\cL_{I_k}(\ahat) = \prod_G \prod_b e^{-2 \pi \lambda_k p_G \int_{B_{d2}}^\infty \left(1-\left(1+\ahat P_k G (\kappa^b_k r_k^{\alpha^b_k} N_b)^{-1}\right)^{-N_b} \right) p^b_k(r_k) r_k dr_k }
	   \end{align}
	   where $B_{d2}=\max(H,Q^{sb}_{kj}(r))$ for $k=1$, and $B_{d2}=Q^{sb}_{kj}(r)$ for $k=2$.

\end{Theo}

\emph{Proof:} See Appendix \ref{Proof_EC}.

\begin{Rem}
	We note that the provided analysis and expressions are general. To find the energy coverage probability of the Thomas cluster process and Mat\'ern cluster process, we only need to substitute the corresponding PDFs and CCDFs in Section III for each cluster process in (\ref{EC}).
\end{Rem}

\begin{Rem}
	Since the harvested energy is a linear funcion of the downlink duration $\tau$, the energy coverage is a monotonically increasing function of $\tau$. On the other hand, the energy coverage probability is monotonically decreasing function of the power splitting parameter $\rho$.
\end{Rem}

\subsection{SINR Coverage Probability}
The SINR coverage probability is defined as the probability that the received SINR is larger than a certain threshold $\gamma_{sinr}>0$. Therefore, given the event $S_{j,s}$, the conditional SINR coverage probability of each tier can be determined using \cite[Theorem 1]{Milli_XWang_journal} and expressed as follows:
\begin{align}\label{CP}
			& P^c_{SINR_{j,s}} (\rho,\tau,\gamma_{sinr})= \prob(\sinr_{j,s}>\gamma_{sinr}|S_{j,s})= \notag \\
			&
			\begin{cases}
				 \sum\limits_{n=1}^{N_s} (-1)^{n+1} \Big(\substack{N_s \\ \\n}\Big)
				 \int\limits_{H}^{\infty} \fhat_{R^s_0}(r) e^{-\mu^s_{j}\left(\frac{\sigma_c^2}{\rho} + \sigma^2_n \right)  }
				 \prod\limits_k \cL_{I_k}(\mu^s_j)dr
				 \hspace{1.62in} \text{for } j=0,\\
				 \sum\limits_{n=1}^{N_s} (-1)^{n+1} \Big(\substack{N_s \\ \\n}\Big)
				 \int\limits_{B_{d1}}^{\infty} \fhat_{R^s_j}(r)
				 e^{-\mu^s_{j}\left(\frac{\sigma_c^2}{\rho} + \sigma^2_n \right)}
				 \cL_{I_0}(\mu^s_j) \prod\limits_k \cL_{I_k}(\mu^s_j) dr \hspace{1in} \text{for }  j=1,2,
			\end{cases}
\end{align}
where $\mu^s_{j}=\frac{n \eta_s \gamma_{sinr} \kappa^s_j r^{\alpha^s_j}}{P_j G_0}$, $\eta_s=N_s(N_s!)^{-\frac{1}{N_s}}$, $N_s$ is the Nakagami fading parameter.
%, and
%\begin{align}
%            & \gamma^G_{k,b}(r) =\left(1+\mu^s_{j} P_k G (\kappa^b_k x^{\alpha^b_k} N_b)^{-1}\right)^{-N_b}
%%			& \label{Lambda} \Lambda^{bG}_s(r)= 2 \pi \lambda_k\\ \notag \\
%%			&\int\limits_{r^{C_{s,b}}}^{\infty} \left(1-\frac{1}{\left(1+\mu_{s} P G \kappa_b x^{-\alpha_b} N_b^{-1}\right)^{N_b}} \right) p_G p_b(x) x dx \\
%%			& \label{gamma}\gamma^{bG}_s(r) = \int\limits_{r^{C_{s,b}}}^{\infty} \frac{p_G D_b  f_{R_0}(x) dx}{\left(1+\mu_{s} P G \kappa_b x^{-\alpha_b} N_b^{-1}\right)^{N_b}F_{R_0}\left( r^{C_{s,b}}\right)}
%\end{align}
\begin{Rem}
	From the downlink SINR expression, we can conclude that the SINR coverage probability is independent of $\tau$. On the other hand, it is a monotonically increasing function of $\rho$.
\end{Rem}

\subsection{Successful Transmission Probability}
In general, the transmission is successful if the UE can both harvest enough energy to charge itself and has sufficient SINR levels for information decoding. Therefore, we define the successful transmission probability (STP) as follows.
\begin{Def}
	       Given that the typical UE is associated with a LOS/NLOS BS from the $j^{th}$ tier, the conditional successful transmission probability is defined as
	       \begin{align}
	       P^c_{ST_{j,s}} (\rho, \tau,\gamma_E, \gamma_{sinr}) = \prob \left(E^{hv}_{j,s} > \gamma_E,  \sinr_{j,s} > \gamma_{sinr}  \right| S_{j,s}).
	       \end{align}
	       Therefore, the total STP of the UAV-assisted mmWave network can be expressed as
	       \begin{align}			
	       & P_{ST} (\rho, \tau,\gamma_E, \gamma_{sinr})= \sum\limits_{j\in\Ksp} P_{ST_{j}} (\rho, \tau,\gamma_E, \gamma_{sinr}) = \sum\limits_{j\in\Ksp} \sum\limits_{s} P^c_{ST_{j,s}} (\rho, \tau,\gamma_E, \gamma_{sinr}) A_{j,s}.			
	       \end{align}
\end{Def}

\begin{Theo}
	Given that the typical UE is associated with a LOS/NLOS BS from the $j^{th}$ tier, the conditional successful transmission probability of each tier can be expressed as
	\begin{align} \label{ST}
			& P^c_{ST_{j,s}} (\rho, \tau,\gamma_E, \gamma_{sinr})=  P^c_{E_{j,s}} (\rho, \tau, \gamma_E)(1-\Fhat_{I}(\omega))  + P^c_{SINR_{j,s}} (\rho, \tau,\gamma_{sinr}) \Fhat_{I}(\omega)
	\end{align}
	where $\omega = \frac{1}{1+\gamma_{sinr}} \left( \frac{\gamma_E}{\tau(1-\rho)}-\gamma_{sinr}\left(\frac{\sigma_c^2}{\rho}+\sigma_n^2\right)\right)$, $P^c_{E_{j,s}} (\rho, \tau,\gamma_E)$ is the conditional energy coverage probability given in (\ref{EC}), $P^c_{SINR _{j,s}} (\rho, \tau,\gamma_{sinr})$ is the conditional SINR coverage probability given in (\ref{CP}), and $\Fhat_{I}(x)$ is the CCDF of $I$ given event $S_{j,s}$, whose expression is as follows:
\begin{align} \label{ICCDF}
		\Fhat_{I}(x)=
			\begin{cases}
				\sum\limits_{n=0}^N (-1)^n \Big(\substack{N \\ \\n}\Big) \int\limits_{H}^{\infty}\fhat_{R^s_0}(r) \prod\limits_k \cL_{I_k}(\ahat') dr
					\hspace{2in} \text{for } j=0,\\
						\sum\limits_{n=0}^N (-1)^n \Big(\substack{N \\ \\n}\Big)
				\int\limits_{B_{d1}}^{\infty}\fhat_{R^s_j}(r)
					\cL_{I_0}(\ahat') \prod\limits_k \cL_{I_k}(\ahat') dr \hspace{1.4in} \text{for } j=1,2,
		   \end{cases}
\end{align}
where $\ahat'=\frac{a n }{x}$.

\end{Theo}

\emph{Proof:} See Appendix \ref{Proof_STP}.

%\begin{Rem}
%	When the noise-limited network is considered, then the STP specializes into $P^c_{ST_{j,s}} (\rho, \tau,\gamma_E, \gamma_{sinr})=  P^c_{E_{j,s}} (\rho, \tau,\gamma_E) \indictor \left( F(\rho, \tau,\gamma_E, \gamma_{sinr}) \geq 0 \right)  +  P^c_{SINR_{j,s}} \indictor \left( F(\rho, \tau,\gamma_E, \gamma_{sinr}) < 0 \right) $, where $F(\rho, \tau,\gamma_E, \gamma_{sinr})=\frac{\gamma_E}{\tau(1-\rho)}- \gamma_{sinr}\left(\frac{\sigma_c^2}{\rho}+\sigma_n^2\right)$. Therefore, depending on the of $\rho$, $\tau$, $\gamma_E$, $\gamma_{sinr}$, STP becomes equal to either the energy coverage or the SINR coverage probability.
%\end{Rem}

\section{Uplink Coverage analysis}
In the uplink phase, UEs use the energy harvested in the downlink phase to transmit data to the cluster center UAVs. We assume all UEs transmit at the fixed power level of $P^{UL}_t$. Then, for successful uplink transmission, the harvested energy $E^{hv}$ for a UE should satisfy
\begin{align}
		E^{hv} \geq (T-\tau) P^{UL}_t.
\end{align}
If this condition is not satisfied, then the UE is assumed to be in inactive mode in the uplink phase, i.e. the UE is not able to transmit; otherwise the UE is in active mode.
Therefore, we can obtain the probability that the UE is in active mode from the energy coverage probability derived in the previous section as follows:
\begin{align}
		p_{active}=P_{E} \left( (T-\tau) P^{UL}_t \right).
\end{align}

\subsection{Uplink SINR coverage}
A UAV from a random cluster is chosen as the typical BS, and a random active UE from the cluster is selected to be the transmitting UE. Note that the active UEs from other clusters will cause interference.
%and the union of these active UEs form a PPP, denoted as $\Phi_{user}$.
Since the links between the typical UAV and the interfering UEs can also be LOS or NLOS, and at most one UE from one cluster inflicts interference, UE can be divided into groups of UEs with LOS and NLOS links, and these groups form PPPs $\Phi^L_{user}$ and $\Phi^N_{user}$ with densities $\lambda^L_{user}=p_{active}p^L_U \lambda_U$ and $\lambda^N_{user}=p_{active}p^N_U \lambda_U$, respectively.  Therefore, the experienced SINR at the typical UAV can be expressed as
\begin{align}
		\sinr^{UL} = \frac{P^{UL}_t G_0 h_0 (k_U^s r_0^{\alpha^s_U})^{-1}}{\sigma_n^2+\sum\limits_{b}\sum\limits_{i\in \Phi^b_{user}} P^{UL}_t G_i h_i (\kappa^b_U r_{i}^{\alpha^b_U})^{-1}}.
\end{align}
where $b\in \{\los,\nlos \}$.
The uplink SINR coverage probability, given the serving UE is in active mode, can be expressed as
\begin{align}
		P^{UL}_{SINR} (\gamma^{UL})=\prob (\sinr^{UL} \geq \gamma^{UL}| active).
\end{align}
\begin{Theo}
		Given that the serving UE is in active mode, then the uplink SINR coverage probability of the network can by expressed as
		\begin{align}
%				P^{UL}_{SINR} (\gamma_{UL})=\E \left[ \sum_{n=1}^{N_s} (-1)^{n+1} \binom{N_s}{n} e^{ - \mu^{UL}_s \sigma_n^2 } \cL_{ I^L_{user} }( \mu^{UL}_s) \cL_{ I^N_{user} }( \mu^{UL}_s) \right]
				P^{UL}_{SINR} (\gamma_{UL})= \sum_{n=0}^{N_s} (-1)^{n+1} \binom{N_s}{n} \int_{H}^{\infty}e^{ - \mu^{UL}_s \sigma_n^2 } \cL_{ I^L_{user} }( \mu^{UL}_s) \cL_{ I^N_{user} }( \mu^{UL}_s) f_{R^s_0} (r_0) dr_0
		\end{align}
		 where $\mu^{UL}_s=\frac{n \eta_s \gamma^{UL}r_0^{\alpha^s_U} }{P^{UL}_t G_0 k_U^s }$. $\cL_{ I^b_{user} }( \mu^{UL}_s)$ is the Laplace transform expression which can be expressed as follows:
		 \begin{align}
		 		\cL_{ I^b_{user} }( \mu^{UL}_s)= \prod_G e^ {-2 \pi p_G  \lambda^b_{user} \int_{0}^{\infty}\left(  1- \int_{0}^{\infty} \left( 1+\mu^{UL}_s  P^{UL}_t G_i (\kappa^b_U (v^2+H^2)^{\frac{\alpha^b_U}{2}} N_b)^{-1} \right)^{-N_b}  f(v|w)dv \right)  wdw }
		 		%\exp\left(-2 \pi\lambda^b_{user} \int_{H}^{\infty}\frac{1}{ 1+(\mu^{UL}_s  P^{UL}_t G)^{-1} \kappa^b_U r^{\alpha^b_U} } rdr \right)
		 \end{align}

\end{Theo}

\emph{Proof:} See Appendix \ref{Proof_UplinkSINR}

\begin{Rem}
	If the small-sale fading $h_i$ of the interfering links are assumed to be Rayleigh distributed, i.e., $N_b=1$, by utilizing the Rician property  $\int_{0}^{\infty} f(v|w) w dw =v$ (when Thomas cluster processes are considered), the Laplace transform can be expressed as
	\begin{align}
		\cL_{ I^b_{user} }( \mu^{UL}_s)= \prod_G\exp \left(-2 \pi \int_{0}^{\infty}\frac{p_G  \lambda^b_{user}}{ 1+(\mu^{UL}_s  P^{UL}_t G)^{-1} \kappa^b_U (v^2+H^2)^{\frac{\alpha^b_U}{2}} } vdv \right) .
	\end{align}	
\end{Rem}

\subsection{Average Throughput}
%Rate of the network can be expressed as
%\begin{align}
%		R^{UL}=(1-\tau) \log_2 (1+\sinr^{UL})
%\end{align}
%Rate coverage can be formulated as $R^{UL}\geq R_{\min}$, and we can re-formulate this by the uplink SINR coverage probability as
%\begin{align}
%		P^{UL}_R=P^{UL}_{SINR}\left( 2^{\frac{R_{min}}{1-\tau}}-1\right)
%\end{align}
The average uplink throughput of the network can be expressed as
\begin{align}
	R^{UL}&=\E\left[(T-\tau)W\log(1+\gamma^{UL})\indictor(\sinr^{UL}\geq \gamma^{UL} )p_{active} \right] \notag \\
	&=(T-\tau)W\log(1+\gamma^{UL}) P^{UL}_{SINR}(\gamma^{UL}) p_{active}
\end{align}
where $W$ is the bandwidth of each channel.
It is also worth noting that $p_{active}$ is related to the energy coverage probability in the downlink phase, and therefore the average uplink throughput has dependence also on the downlink phase. With this, we formulate the following optimization problem to maximize $R^{UL}$ subject to a lower bound constraint on the downlink throughput
\begin{align}\label{Rate_Opt}
	& \max_{\tau} \quad (T-\tau)W\log(1+\gamma^{UL}) P^{UL}_{SINR}(\gamma^{UL}) p_{active} \notag \\
	&  s.t. \qquad R^{DL} \geq R_{\min}
\end{align}
where $R^{DL}= \tau W\log(1+\gamma^{UL}) P_{SINR}(\gamma_{sinr})$ is the average donwlink throughput, $R_{\min}$ is the minimum average throughput requirement for the downlink transmission. We numerically solve this problem in the Section VIII.

\section{Generalizations and Special Cases }
While we have assumed in the previous sections that the UAVs fly at the same height, our analysis and approach are relatively broad. To demonstrate this, we extend our analysis to a multi-tier multi-height model in this section. Additionally, we address the special case of the noise-limited network and  derive closed-form characterizations with practical implications.

\subsection{Multi-tier multi-height model}
In practice, UAVs can fly at different heights depending on the applications and regulations.  For instance, UAV heights may differ in urban areas with high-rise buildings compared to suburban environments.  With this motivation, we consider a multi-tier multi-height model, in which we have multiple tiers of UAVs and UAVs in the $j^{th}$ tier are distinguished with their density $\lambda_j$, transmit power $P_j$, biasing factor $B_j$ and height $H_j$. Next, we discuss how our previous analysis can be adapted to this model.

Suppose we have $\Ksp_U =\{1,2,...,K \}$ tiers of UAVs. Then we introduce two notations: $\Ksp_G = \{GBS\}  \cup \Ksp_U  $ and $\Ksp= \{0\} \cup \{GBS\}  \cup \Ksp_U  $. Since we still use the same downlink association criterion, the received power can be re-expressed as
\begin{align}
P_m&=  \argmin\limits_{j \in \Ksp, s } P_j G_0 B_j (\kappa^s_j (r^s_j)^{\alpha^s_j}).
%=\argmin_{j \in \{0,1,2\}, i\in \Phi} \kappa_{}
\end{align}
Now, the probability that the typical UE is associated with a LOS/NLOS BS in the $j^{th}$ tier can be modified from (\ref{AP}) as
	\begin{align}
	A_{j,s}=
	\begin{cases}
	\E_{R^s_{0}} \left[  D_0^s \prod\limits_{k \in \Ksp_G} \prod\limits_{b} D_k^b  \Fbar_{R_k^b} \left(Q_{k0}^{sb} (r_0)\right)\right],  \hspace{2.9in}\text{for }j=0, \\
	D_j^s\E_{R^s_{j}}  \Bigg[D_j^{s'} \Fbar_{R^{s'}_{j}}\left(Q_{jj}^{ss'} (r_j)\right)
	\hspace{0in} \left(\sum\limits_b D_0^b \Fbar_{R^b_{0}}\left(Q_{0j}^{sb}(r_j) \right)\right) \prod\limits_{\substack{k \in \Ksp_G \\ k \neq j}} \prod\limits_b D_k^b  \Fbar_{R_k^b} \left(Q_{kj}^{sb} (r_j)\right) \Bigg],  \hspace{0.10in} \text{for }j \in \Ksp_1 .
	%    \hspace{2.10in} \text{for }j=1,2,
	\end{cases}
	\end{align}
	Similarly, the energy coverage, SINR coverage and the successful transmission probabilities can be modified from (\ref{EC}), (\ref{CP}) and (\ref{ST}), respectively, by letting $k \in \Ksp_G$. The CCDFs and PDFs of the distances remain the same.

\subsection{Noise-limited model}
In this subsection, we investigate the network performance metrics when the interference is negligible. When interference $I \approx 0$, the energy coverage and SINR coverage probabilities can be simplified by removing the Laplace transform terms in   (\ref{EC})  and (\ref{CP}), respectively. 	With this, the STP specializes to
\begin{align}
		P^c_{ST_{j,s}} (\rho, \tau,\gamma_E, \gamma_{sinr})=  P^c_{E_{j,s}} (\rho, \tau,\gamma_E) \indictor \left( F(\rho, \tau,\gamma_E, \gamma_{sinr}) \geq 0 \right)  +  P^c_{SINR_{j,s}} \indictor \left( F(\rho, \tau,\gamma_E, \gamma_{sinr}) < 0 \right)
\end{align}
where $F(\rho, \tau,\gamma_E, \gamma_{sinr})=\frac{\gamma_E}{\tau(1-\rho)}- \gamma_{sinr}\left(\frac{\sigma_c^2}{\rho}+\sigma_n^2\right)$. The partial derivative of $F$ with respect to $\rho$ can be expressed as
\begin{align}
	\frac{\partial F}{\partial \rho} = \frac{\gamma_E}{\tau (1-\rho)^2} + \frac{\gamma_{sinr} \sigma_c^2}{\rho^2} > 0.
\end{align}
Hence $F$ is a monotonically increasing function of $\rho$. Therefore, depending on the values of $\tau$, $\gamma_E$ and $ \gamma_{sinr}$, there are three cases: 1) if $F_{\max} <0$, $P^c_{ST_{j,s}} = P^c_{SINR_{j,s}}$; 2) if $F_{\min} >0$,  $P^c_{ST_{j,s}} = P^c_{E_{j,s}}$; 3) if  $F_{\max} >0$ and $F_{\min} <0$, then in region of $F<0$ we have $P^c_{ST_{j,s}} = P^c_{SINR_{j,s}}$, which is a monotonically increasing function of $\rho$, while in region of $F>0$ we have $P^c_{ST_{j,s}} = P^c_{E_{j,s}}$, which is a monotonically decreasing function of $\rho$, therefore, with increasing $\rho$, $P^c_{ST_{j,s}}$ first increases then decreases,  and  $F = 0$  gives the maximum of $P^c_{ST_{j,s}}$, i.e. when
\begin{align}\label{rho}
	\rho^* = \frac{-(\gamma_E+\tau \gamma_{sinr} \sigma_c^2 - \tau \gamma_{sinr} \sigma_n^2)+\sqrt{(\gamma_E+\tau \gamma_{sinr} \sigma_c^2 - \tau \gamma_{sinr} \sigma_n^2)^2-4\tau^2\gamma_{sinr}^2 \sigma_c^2 \sigma_n^2}}{2 \tau \gamma_{sinr} \sigma_n^2}.
\end{align}
%
% $\rho = \frac{\sqrt{(\gamma_E+\tau \gamma_{sinr} \sigma_c^2 - \tau \gamma_{sinr} \sigma_n^2)^2-4\tau^2\gamma_{sinr}^2 \sigma_c^2 \sigma_n^2}}{2 \tau \gamma_{sinr} \sigma_n^2}$.

When we further assume that the uplink between the typical UAV and its cluster member UE is in LOS, and the path-loss exponent is $\alpha^L_U = 2$ and the small-scale fading is Rayleigh fading, the uplink SNR coverage probability admits the following simpler expression:
\begin{align}
	P^{UL}_{SINR} (\gamma^{UL}) =
	\begin{cases}
		% \sum_{n=0}^{N_s} (-1)^{n+1} \binom{N_s}{n}
		\frac{e^{-C'H^2}}{1+ 2 C' \sigma^2}  \hspace{1.1in} \text{for Thomas cluster process,}\\
		 %\sum_{n=0}^{N_s} (-1)^{n+1} \binom{N_s}{n}
		  \frac{e^{-C'H^2}}{C'R_c^2} \left(1-e^{-C'R_c^2}\right) \hspace{0.2in} \text{for Mat\'ern cluster process,}
	\end{cases}
\end{align}
where $C'= \frac{ \gamma^{UL} \sigma_n^2 }{P^{UL}_t G_0 k_U^L }$.
 % $C'= \frac{n \eta_L \gamma^{UL} \sigma_n^2 }{P^{UL}_t G_0 k_U^L }$.

\section{Numerical Results}
In this section, we provide numerical results to evaluate the performance of the considered UAV-assisted mmWave cellular network and identify the impact of key network parameters on the performance.
Unless stated otherwise, the parameter values are listed in Table \ref{table_values}.
% $P_u=24$ dBm \cite{UAV_CLiu}, $P_G=34$ dBm, $P_t^{UL}=1$ dBm, $H=50$ m, $\lambda_U=10^{-4} $ /m$^2$ and $\lambda_G=10^{-5} $ /m$^2$. For urban environment, the path loss parameters are $C=11.95$, $B=0.136$, $\kappa^L_j=10^{3.08}$, $\kappa^N_j=10^{0.27}$, $\alpha^L_j=2.09$, $\alpha^N_j=3.75$ \cite{UAV_CLiu}\cite{UAV_MMozaffari} and $1/\beta=141.4$ \cite{Milli_Bai}\cite{EH_TKhan}. Following \cite{Milli_Bai}\cite{EH_TKhan}, the mmWave carrier frequency is assumed to be 28 GHz, the bandwidth $W=100$ MHz, $\sigma_c^2= -80$ dB and $\sigma_n^2=$-174dBm/Hz+10$\log_{10}$($W$)+10dB. The parameters for Nakagami fading are $N_L=2$ and $N_N=3$. The total time duration $T=1$s.

\begin{table}[htbp]
	\caption{Table of Parameter Values}
	\centering
	\begin{tabular}{|l|p{2.5in}|}
		\hline
		\footnotesize \textbf{Notations} &  \footnotesize \textbf{Description} \\ \hline		
		\scriptsize  $P_u, P_G, P_t^{UL}$ &  \scriptsize  24 dBm \cite{UAV_CLiu}, 34 dBm, 1 dBm \\\hline
		\scriptsize $\lambda_U, \lambda_G $   &  \scriptsize $10^{-4} $ /m$^2$, $10^{-5} $ /m$^2$ \\\hline
		\scriptsize $H, C, B$  &  \scriptsize 50 m, 11.95, 0.136 \cite{UAV_CLiu}\cite{UAV_MMozaffari} \\\hline
		\scriptsize  $\kappa^L_j,\kappa^N_j,\alpha^L_j, \alpha^N_j$   &  \scriptsize  $10^{3.08}$, $10^{0.27}$, 2.09, 3.75 \cite{UAV_CLiu}\cite{UAV_MMozaffari} \\\hline
		\scriptsize $1/\beta$  &  \scriptsize 141.4 \cite{Milli_Bai}\cite{EH_TKhan} \\\hline
		\scriptsize Carrier frequency, $W$  &\scriptsize   28 GHz, 100 MHz \cite{Milli_Bai}\cite{EH_TKhan}\scriptsize  \\\hline
		\scriptsize $\sigma_n^2, \sigma_c^2$  &\scriptsize -174 dBm/Hz+10$\log_{10}$($W$)+10 dB, -80 dB \cite{Milli_Bai}\cite{EH_TKhan} \scriptsize  \\\hline
		\scriptsize $N_L, N_N$  &\scriptsize 2, 3 \scriptsize  \\\hline		
		\scriptsize $T$  &\scriptsize 1 s \scriptsize  \\\hline		
	\end{tabular}
    \label{table_values}
\end{table}\normalsize

\subsection{Impact of the cluster size}
\begin{figure}
	\centering
	\begin{minipage}{0.45\textwidth}
		\centering
		\includegraphics[width=1\textwidth]{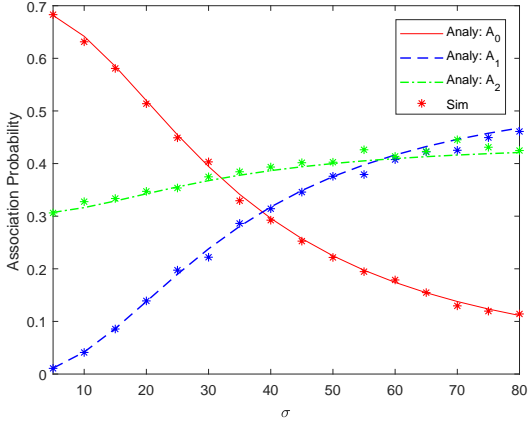} \\
		\subcaption{\scriptsize Thomas cluster process. }
	\end{minipage}
	\begin{minipage}{0.45\textwidth}
		\centering
		\includegraphics[width=1\textwidth]{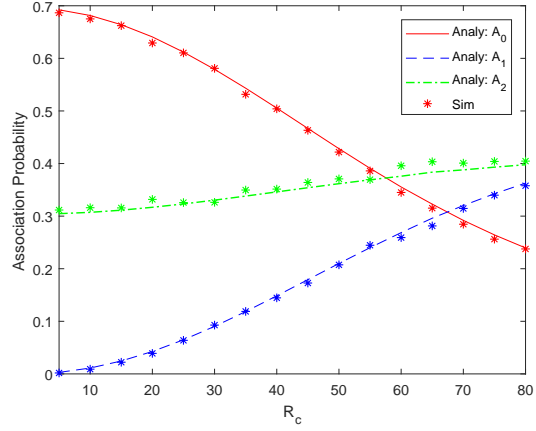}
		\subcaption{\scriptsize Mat\'ern cluster process.}
	\end{minipage}
	\caption{\small Association probability as functions of the cluster size with parameter values listed in Table \ref{table_values}.  \normalsize}
	\label{AP_clustersize}
\end{figure}
\begin{figure}
	\centering
	\begin{minipage}{0.45\textwidth}
		\centering
		\includegraphics[width=1\textwidth]{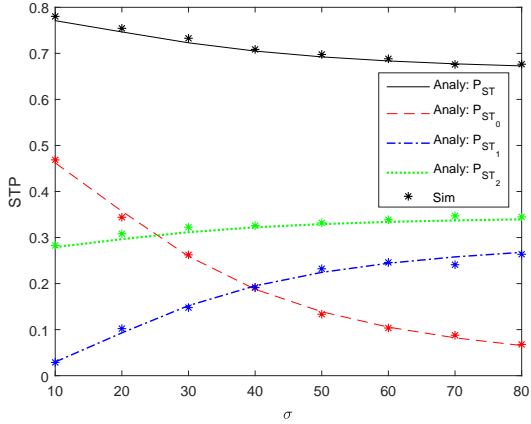} \\
		\subcaption{\scriptsize Thomas cluster process. }
	\end{minipage}
	\begin{minipage}{0.45\textwidth}
		\centering
		\includegraphics[width=1\textwidth]{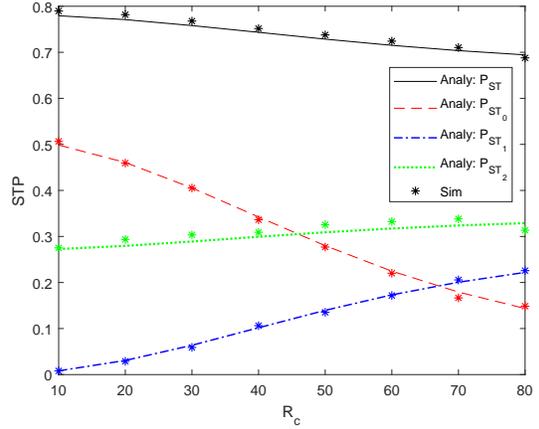}
		\subcaption{\scriptsize Mat\'ern cluster process.}
	\end{minipage}
	\caption{\small STP of the network and each tier BSs as a function of the cluster size when $\tau=T$, $\gamma_E=-40$ dB, $\gamma_{sinr}=0$ dB and $\rho=0.5$. \normalsize}
	\label{STP_clustersize}
\end{figure}
\begin{figure}
	\centering
	\begin{minipage}{0.45\textwidth}
		\centering
		\includegraphics[width=1\textwidth]{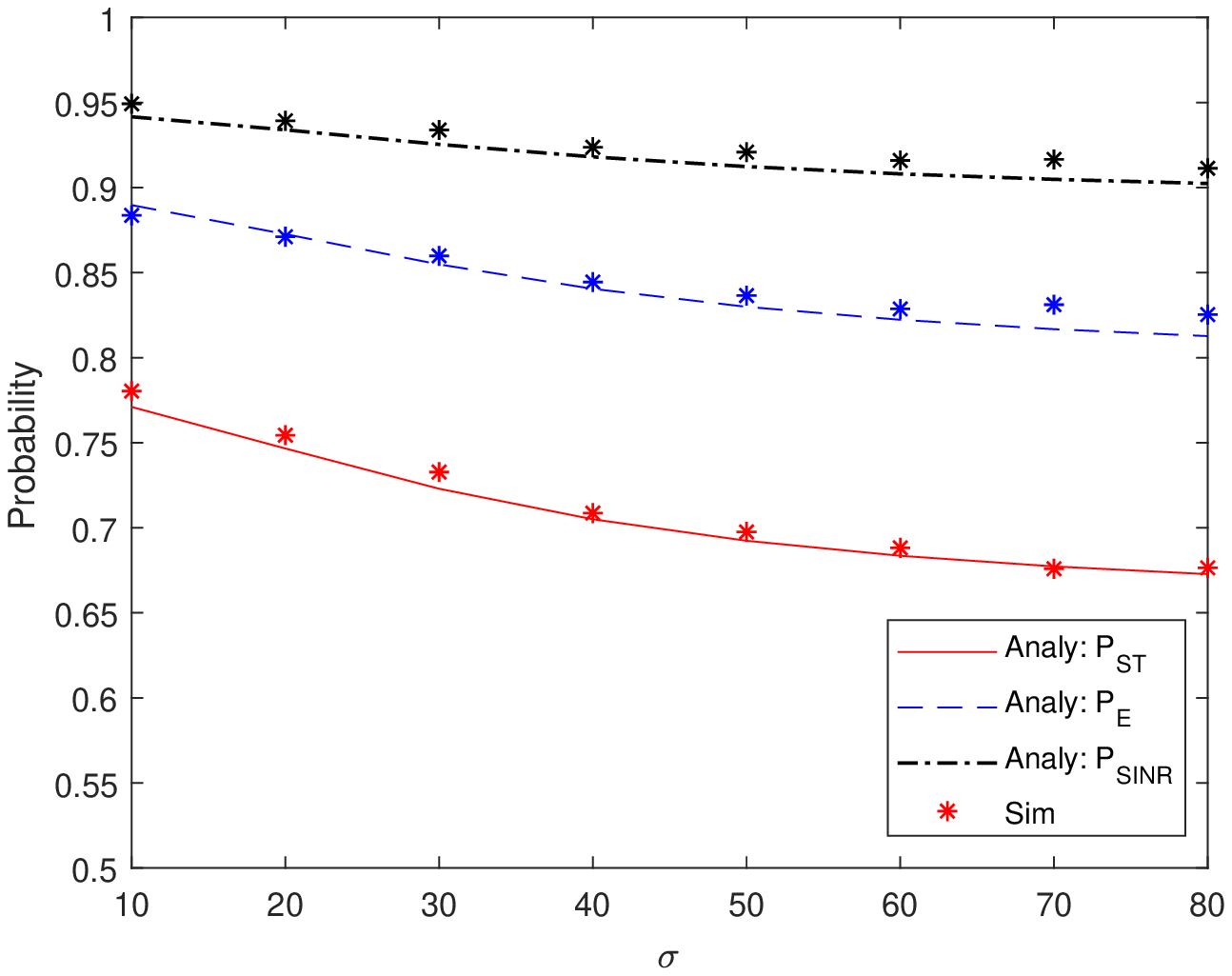} \\
		\subcaption{\scriptsize Thomas cluster process. }
	\end{minipage}
	\begin{minipage}{0.45\textwidth}
		\centering
		\includegraphics[width=1\textwidth]{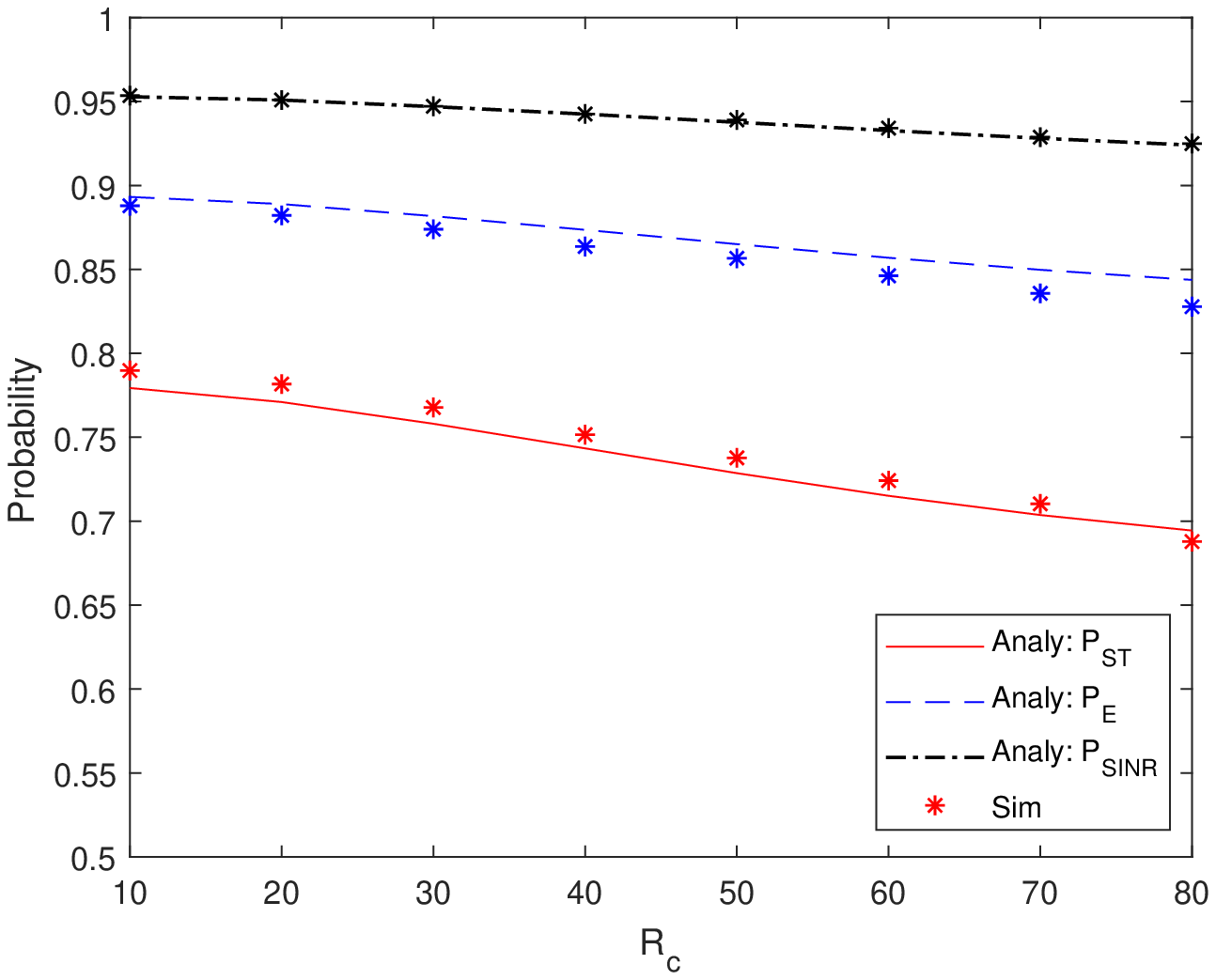}
		\subcaption{\scriptsize Mat\'ern cluster process.}
	\end{minipage}
	\caption{\small STP, EC and SINRC probabilities as functions of the cluster size when $\tau=T$, $\gamma_E=-40$ dB, $\gamma_{sinr}=0$ dB. And $\rho=0.5$ for the SWIPT scenario.  \normalsize}
	\label{Downlink_clustersize}
\end{figure}

 First we investigate the influence of the cluster size on the network performance. The cluster size here is the spatial size of the cluster. More specifically, for the Thomas cluster process, 68.27\% of UEs are located inside a circular region with radius $\sigma$, and 95.45\% of UEs are located insider a circular region with radius $2\sigma$, and we choose $\sigma$ as the cluster size. For Mat\'ern cluster process, $R_c$ is the cluster size.
\subsubsection{Downlink association probability}
 Fig. \ref{AP_clustersize} shows the association probability (AP) as a function of the cluster size in the downlink phase. As shown in the figure, when we increase $\sigma$ and $R_c$, $A_0$ decreases while $A_1$ and $A_2$ increase. As $\sigma$ and $R_c$  increase, the UEs move further away from the projection of the cluster center UAV and hence are more spread away. As a result, the UEs move closer to other UAVs and GBSs. Therefore, $A_0$ decreases. On the other hand, because of the LOS probability function, the link between the UE and the UAVs are more likely to be LOS, and consequently the UEs prefer to be served by UAVs with higher probability. For this reason,  $A_1$ increases faster than $A_2$. We also note that in Fig. \ref{AP_clustersize} (and in the subsequent figures in this section), simulation results are plotted with $*$ markers and we generally observe excellent agreements with the analytical results, further confirming, for instance, our characterizations in Lemma 3 in this case.

\subsubsection{Downlink coverage probabilities}
  Fig. \ref{STP_clustersize} shows the  successful transmission probability (STP) as a function of the cluster size. Since the STP, energy coverage (EC), SINR coverage (SINRC) performances of each tier BS are similar, we evaluate the STP performance of each tier in the figure.
  In this figure, total STP decreases with increasing $\sigma$ and $R_c$. As expected,  when $\sigma$ and $R_c$ become larger, $P_{ST_0}$ (i.e., the successful transition probability in tier 0) diminishes while  $P_{ST_1}$ and  $P_{ST_2}$ increase. However, since the cluster center UAV can provide the maximum conditional coverage, the increase in $P_{ST_1}$ and  $P_{ST_2}$ is not able to compensate the decrease in $P_{ST_0}$, leading to the decrease in total STP.

  In Fig. \ref{Downlink_clustersize}, we observe that STP, EC and SINRC are all monotonically decreasing functions of $\sigma$ and $R_c$. Additionally, we note that since we consider the SWIPT scenario with $\rho=0.5$, we divide the received power of the typical UE into two streams, one for energy harvesting and the other for information decoding. Due to this, the STP performance is lower compared to only EC or SINRC, where it is assumed that entire received power is used for energy harvesting or information decoding only.

\subsubsection{Uplink coverage probability}
\begin{figure}
	\centering
	\begin{minipage}{0.45\textwidth}
		\centering
		\includegraphics[width=1\textwidth]{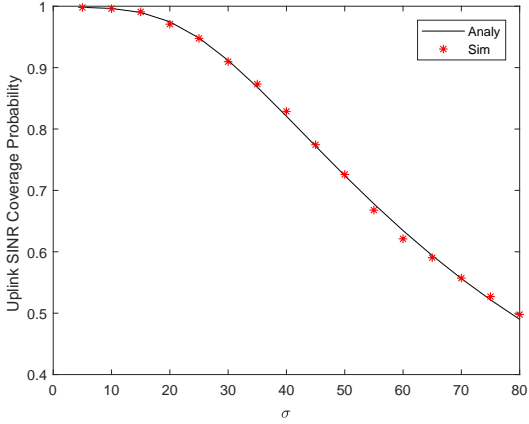} \\
		\subcaption{\scriptsize Thomas cluster process. }
	\end{minipage}
	\begin{minipage}{0.45\textwidth}
		\centering
		\includegraphics[width=1\textwidth]{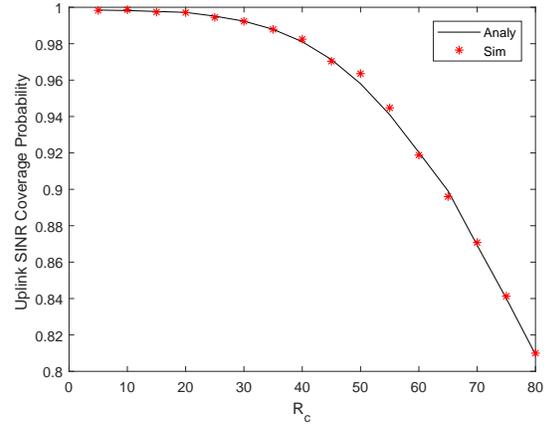}
		\subcaption{\scriptsize Mat\'ern cluster process.}
	\end{minipage}
	\caption{\small Uplink SINR coverage probability as a function of the cluster size when $\tau=0.5T$, $\rho=0$, $\gamma^{UL}=-20$dB, and $\sigma=10$. \normalsize}
	\label{Uplink_clustersize}
\end{figure}
We observe from Fig. \ref{Uplink_clustersize} that the uplink SINRC is a monotonically decreasing function of the cluster size, similarly as in the downlink phase. When compared with the downlink SINRC (blue dashed line) in Fig. \ref{Downlink_clustersize}, we notice in Fig. \ref{Uplink_clustersize} that the uplink SINRC drops faster than the downlink SINRC for larger thresholds. This is due to the different association criteria in different phases. In the downlink phase, because of the strongest long-term averaged received power association criterion, when the UEs are more spread away from their cluster center UAVs, they can get associated with other UAVs and GBSs to get the strongest received power. But in the uplink phase, UAVs are receiving information from their cluster member UEs, and therefore when the UEs are far away, the uplink SINRC  decreases substantially.

Again, we note that simulation results are also provided in all the figures using markers, and these results match with the analytical results, further validating the accuracy of our coverage analysis.  Additionally, we observe in the numerical results that Thomas cluster processes and Mat\'ern cluster processes generally lead to similar network performance trends, which gives us the insight that considering PCP rather than PPP is the key to capture the UE distribution. Therefore, for brevity, we will just provide numerical results considering Thomas cluster processes in the following subsections.

%\subsection{Impact of the interference and performance threshold}
%\begin{figure}
%	\centering
%	\begin{minipage}{0.45\textwidth}
%		\centering
%		\includegraphics[width=1\textwidth]{SWIPT_analy_sim_w_T_sigma10_H50_rho5.eps} \\
%		\subcaption{\scriptsize Successful transmission probability. }
%	\end{minipage}
%	\begin{minipage}{0.45\textwidth}
%		\centering
%		\includegraphics[width=1\textwidth]{All_interference.eps}
%		\subcaption{\scriptsize STP, EC and SINRC performances.}
%	\end{minipage}
%	\caption{\small STP, EC and SINRC as functions of the threshold when $\sigma=10$. \normalsize}
%	\label{figure_interference}
%\end{figure}
%
%Fig. \ref{figure_interference} plots the STP, EC and SINRC as functions of the threshold in the probability formulations. From Fig. \ref{figure_interference}(a), we readily observe that the cluster center UAV provides the largest coverage compared to other UAVs and GBSs. In addition, STP is a monotonically decreasing function of the threshold due to the definition of STP. In Fig. \ref{figure_interference}(b), the STP, EC and SINRC are displayed with and without interference. Due to the fact that the transmit power of UAVs and GBAs are small, interference does not have significant impact on the system performance. At the same time, we can still observe from the figure that interference can lead to an increase in EC and STP, but a decrease in SINRC.

\subsubsection{Multi-tier multi-height model}
\begin{figure}
	\centering
	\begin{minipage}{0.45\textwidth}
		\centering
		\includegraphics[width=1\textwidth]{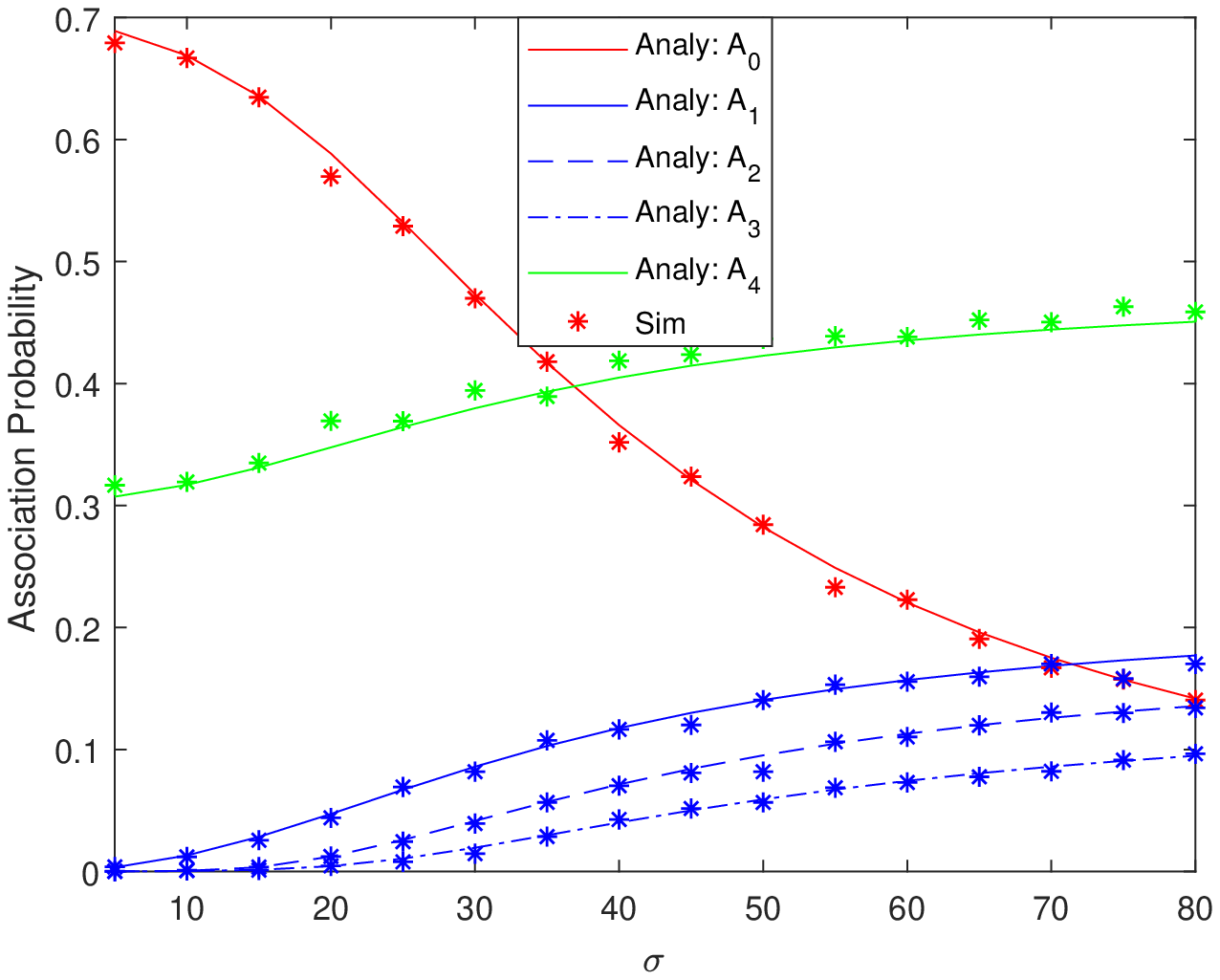}
		\subcaption{\scriptsize Association probability.}
	\end{minipage}
	\begin{minipage}{0.45\textwidth}
		\centering
		\includegraphics[width=1\textwidth]{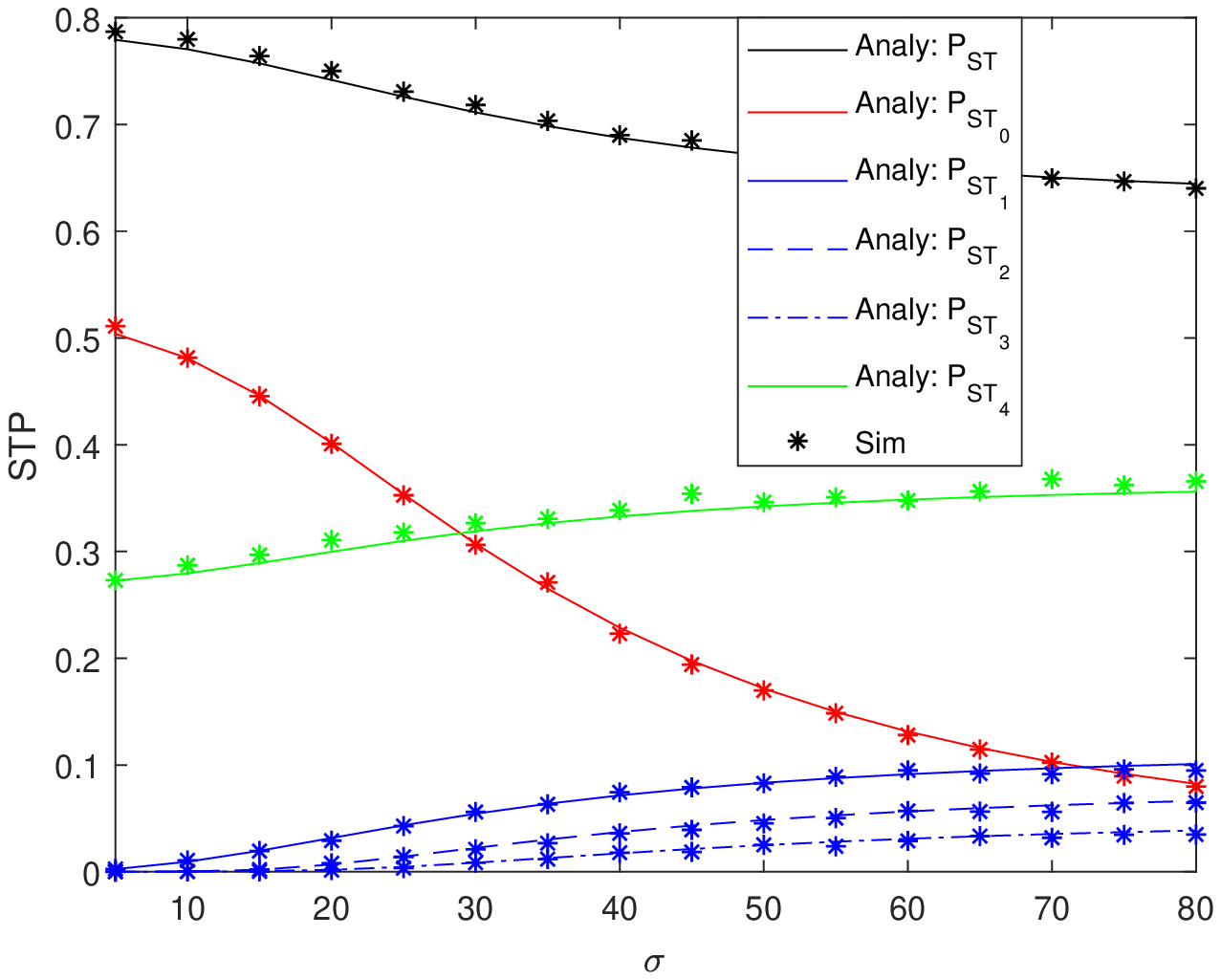}
		\subcaption{\scriptsize Successful transmission probability.}
	\end{minipage}
	\caption{\small Association probability and STP as a function of $\sigma$ for Thomas cluster process, when $\gamma_E = -40$ dB,  $\gamma_{sinr}=0$ dB, $\rho=0.5$ and heights for 0$^{th}$-3$^{rd}$ tier UAVs are 50m, 50m, 60m, and 70m, respectively. The GBS is regarded as the $4^{th}$ tier.\normalsize}
	\label{Figure_multi_H}
\end{figure}

In this part, we assume there are three tiers of UAVs with heights 50m, 60m, and 70m, respectively,  and density $3\times 10^{-5}$/m$^2$. And all UAVs have their own clustered UEs on the ground. There is a tier of GBSs with parameter values listed in Table III. We randomly choose a UE from a cluster of the 50m-high UAVs, and provide the association probability and STP in Figs. \ref{Figure_multi_H}(a) and \ref{Figure_multi_H}(b), respectively. In Fig. \ref{Figure_multi_H}(a), we observe that the association probability of the GBS does not change much when compared with the one-tier UAV model. And as $\sigma$ increases, the association probabilities of $1^{st}$, $2^{nd}$ and $3^{rd}$ tier UAVs increase. Fig. \ref{Figure_multi_H}(b) shows the similar performance levels as in Fig. \ref{STP_clustersize}. And the STP of the  $1^{st}$, $2^{nd}$ and $3^{rd}$ all increase with increasing $\sigma$.

\subsection{Impact of the interference}
\begin{figure}
	\centering
	\includegraphics[width=0.45\textwidth]{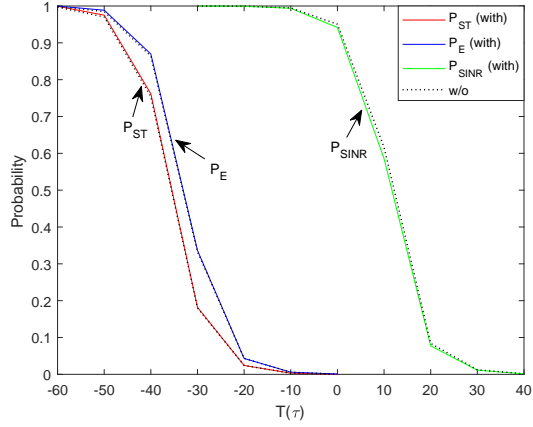} \\
	\caption{\small STP, EC and SINRC as a function of the threshold when $\sigma =10$, $\tau =T$ and $\rho = 0.5$ for the SWIPT scenario. }
	\label{figure_interference}
\end{figure}

In this section, we investigate the impact of the interference. In Fig. \ref{figure_interference}, we plot the EC, coverage probability and STP as a function of the threshold in the downlink phase. Since the GBSs with large transmit power are relatively far from the typical UE and the UAVs which are relatively denser and closer but with low transmit power, the interference is negligible. Thus the interference has little impact on the uplink SINRC. Therefore, as expected the interference does not lead to a significant difference on the probabilities. In the uplink phase, the interference from the UEs is small and has unnoticeable impact on the uplink SINRC.

\subsection{Impact of the UAV height}
\begin{figure}
	\centering
	\includegraphics[width=0.45\textwidth]{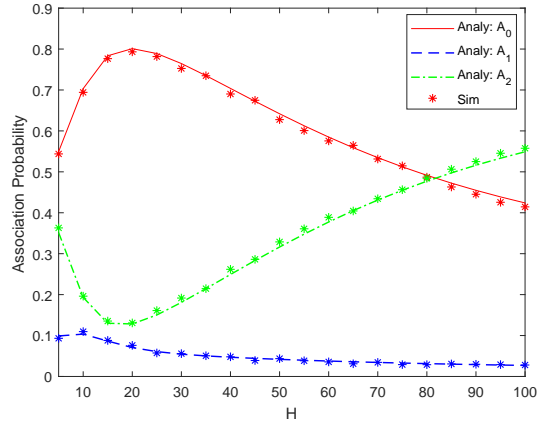} \\
	\caption{\small Association probability as a function of the UAV height with parameter values listed in Table \ref{table_values}. }
	\label{AP_H}
\end{figure}
In this subsection, we investigate the impact of the UAV height on the network performance.
\subsubsection{Downlink association probability}
Fig. \ref{AP_H} depicts the AP of each tier BS as a function of $H$. When $H=0$, the UAVs are located on the ground. Since the UAVs are more densely distributed than the GBSs, we have $A_0>A_2>A_1$. Also because $p^L_U$ is a monotonically increasing function of $H$, the LOS probability of UAVs increases with increasing $H$. Therefore, as $H$ becomes slightly larger, AP with the cluster center UAV, $A_0$,  and AP with other UAVs, $A_1$, increase while AP with GBSs, $A_2$, decreases. On the other hand, when we increase $H$ substantially (e.g., beyond approximately 20m), the UAVs start being high above the sky. Therefore, even though the LOS probabilities have grown, the distances between the UAVs and UEs have increased as well (increasing the path loss), while the distance between the UEs and GBSs have not changed. Due to this, we observe that $A_0$ and $A_1$ decrease whereas $A_2$ starts increasing.

%Therefore as $H$ becomes larger, $A_0+A_1$ increases while $A_2$ decreases. On the other hand, when we slightly increase $H$, the link to the cluster center UAV is LOS with higher probability, while the link to other UAVs is more likely to be NLOS because the signal can be obstructed by buildings. Consequently, $A_0$ increases while $A_1$ decreases. When we increase $H$ substantially, the UAVs are high above the sky and more links from the typical UE to the UAVs become LOS, and hence the UEs can be served by other UAVs, leading to a decreasing $A_0$ and an increasing $A_1$.

\begin{figure}
	\centering
	\begin{minipage}{0.45\textwidth}
		\centering
		\includegraphics[width=1\textwidth]{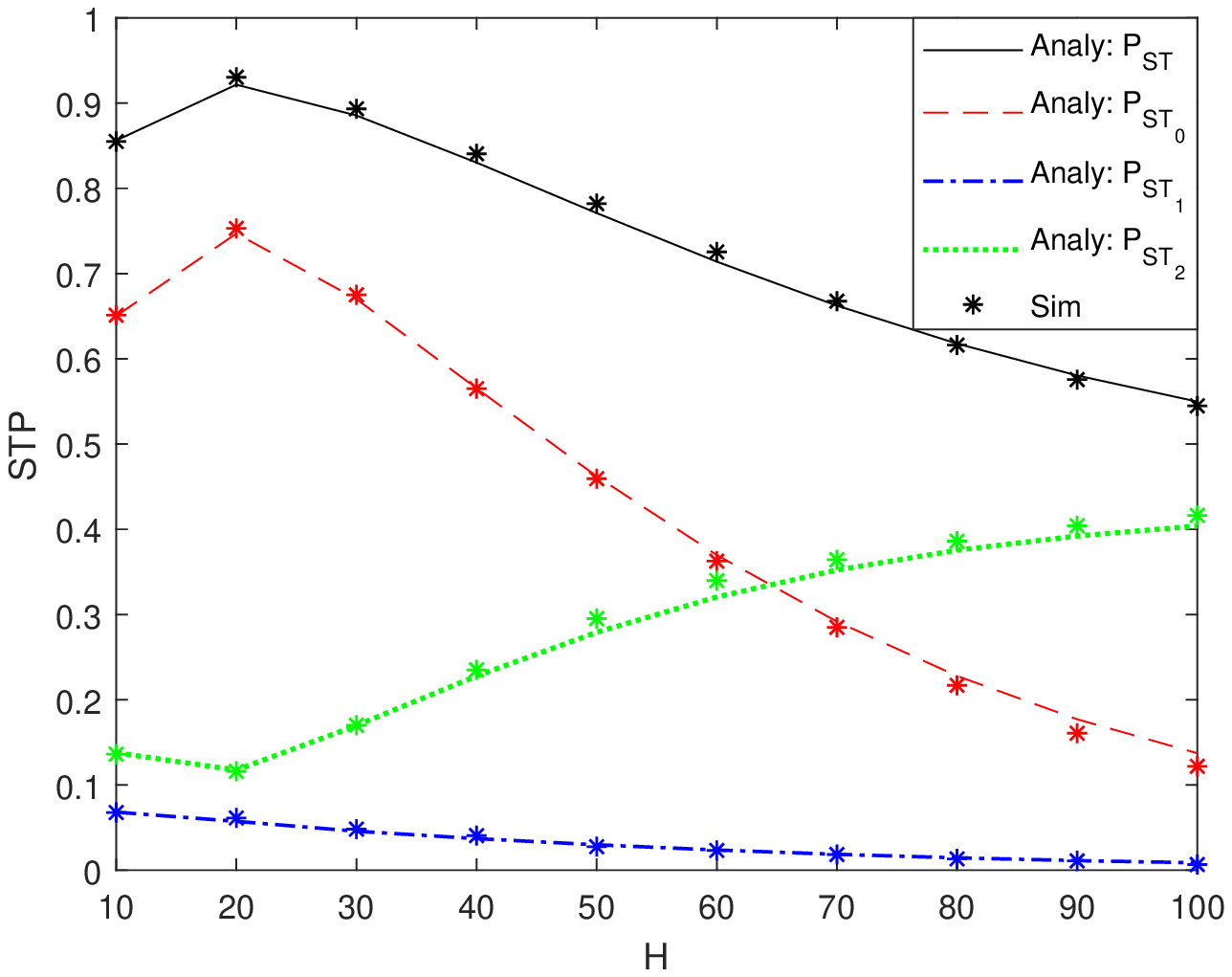}
		\subcaption{\scriptsize Successful transmission probability.}
	\end{minipage}
	\begin{minipage}{0.45\textwidth}
		\centering
		\includegraphics[width=1\textwidth]{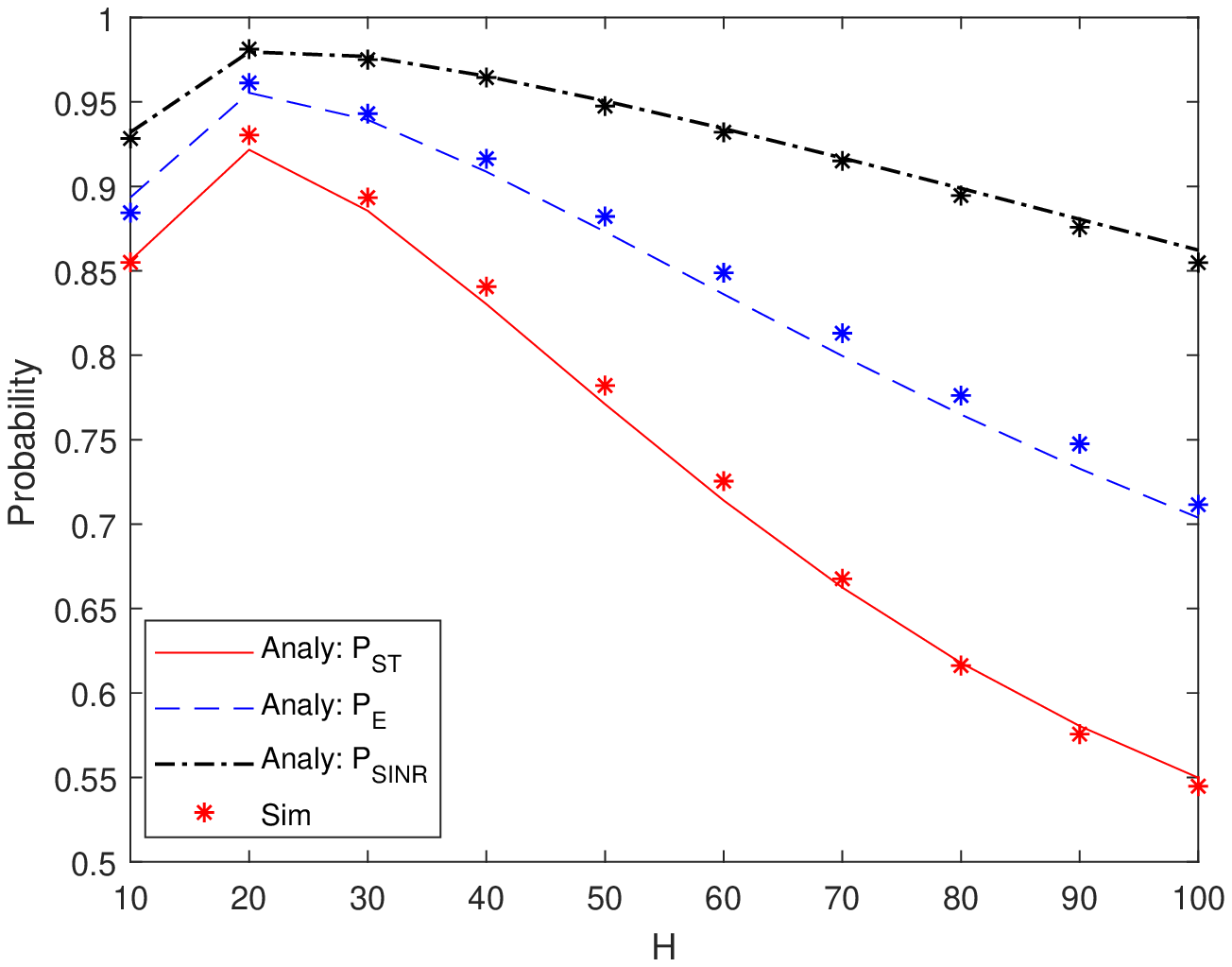}
		\subcaption{\scriptsize STP, EC and SINRC performances.}
	\end{minipage}
	\caption{\small STP, EC and SINRC as functions of the UAV height $H$ when $\sigma=10$, $\tau=T$,  $\gamma_E=-40$ dB, $\gamma_{sinr}=0$ dB. And $\rho=0.5$ for the SWIPT scenario. \normalsize}
	\label{figure_H}
\end{figure}
\subsubsection{Downlink coverage probabilites}
The STP performance curves of each tier BS shown in Fig. \ref{figure_H}(a) demonstrate the same trends as the association probability in Fig. \ref{AP_H}. In addition, the total STP initially grows, achieves its maximum around $H \approx 20$m, and then decreases because of the increased distance between the UEs and UAVs when the UAV height $H$ becomes larger.
At these larger height levels, the increase in $P_{ST_2}$  cannot compensate the decrease in $P_{ST_0}$ and  $P_{ST_1}$. Fig. \ref{figure_H}(b) shows that the EC and SINRC performances follow the same trends as for STP.
\subsubsection{Uplink coverage probability}
\begin{figure}
	\centering
	\includegraphics[width=0.45\textwidth]{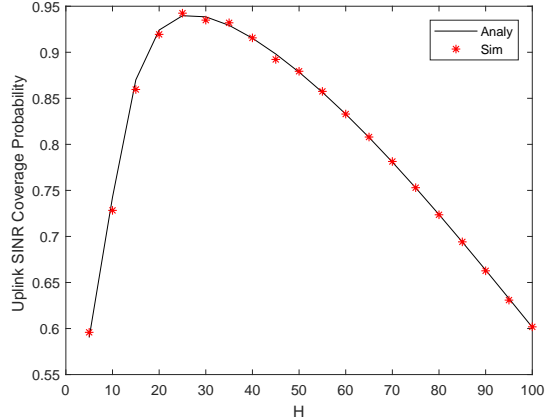} \\
	\caption{\small Uplink SINR coverage probability as a function of the UAV height, when  $\tau=0.5T$, $\rho=0$, $\gamma^{UL}=-20$ dB, and $\sigma=10$. }
	\label{Uplink_H}
\end{figure}
In the uplink phase, the UAVs are receiving data from their cluster member UEs. When UAVs are at relatively lower height, the transmission are NLOS with high probability because of the blockage from buildings and other large objects.  Since the blockage becomes less when we increase the UAV height, the SINRC increases. However, above a certain height, the distance between the UAV and the serving UEs becomes large enough that the path loss starts dominating and as a result, SINRC diminishes. Therefore, as shown in Fig. \ref{Uplink_H}, SINRC increases at first and then decreases, and there exists an optimal height, which is not the same but very close to the optimal height in the downlink phase.

\subsection{Impact of the power splitting component}
\begin{figure}
	\centering
	\begin{minipage}{0.32\textwidth}
		\centering
		\includegraphics[width=1\textwidth]{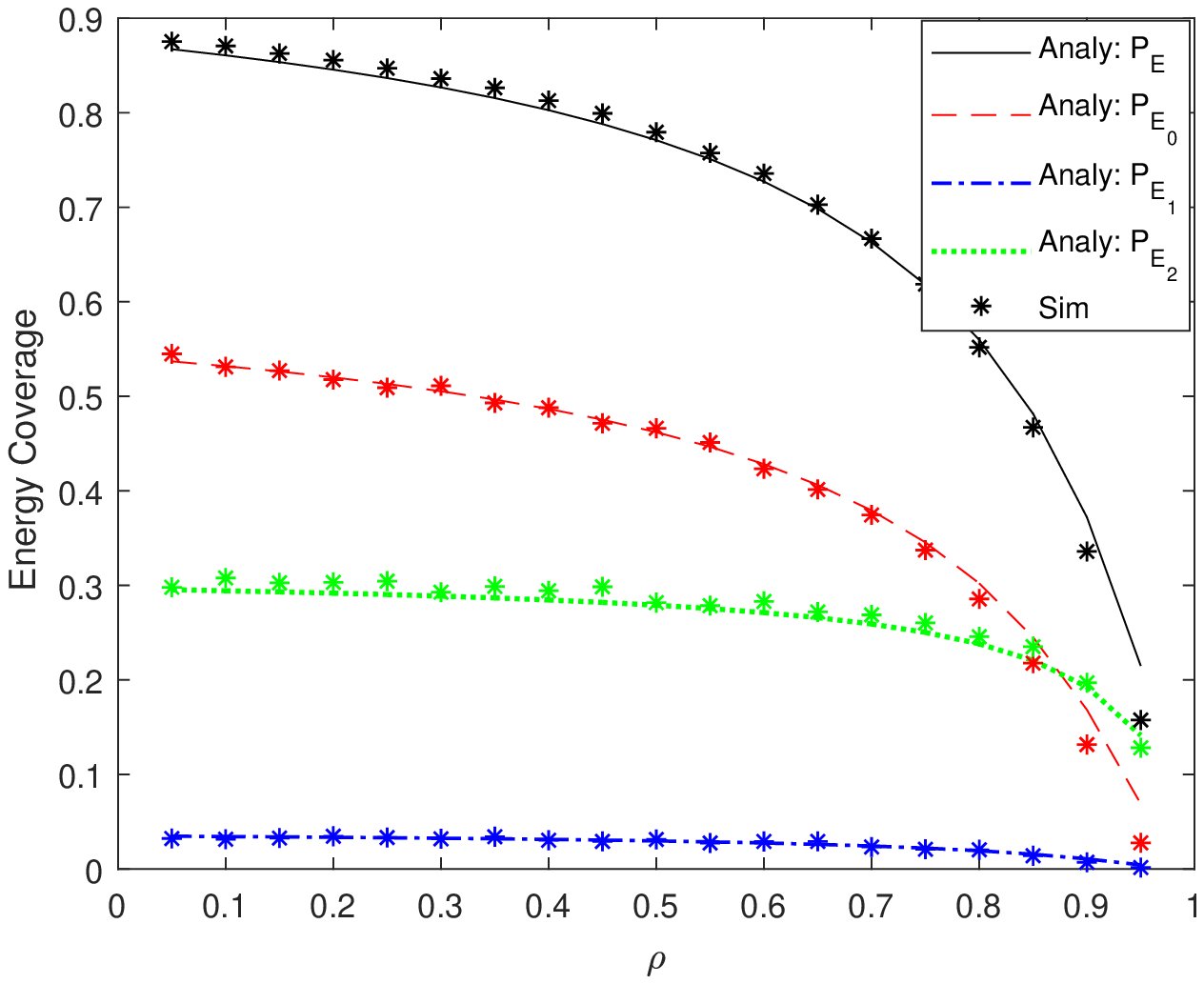}
		\subcaption{\scriptsize Energy coverage. }
	\end{minipage}
	\begin{minipage}{0.32\textwidth}
		\centering
		\includegraphics[width=1\textwidth]{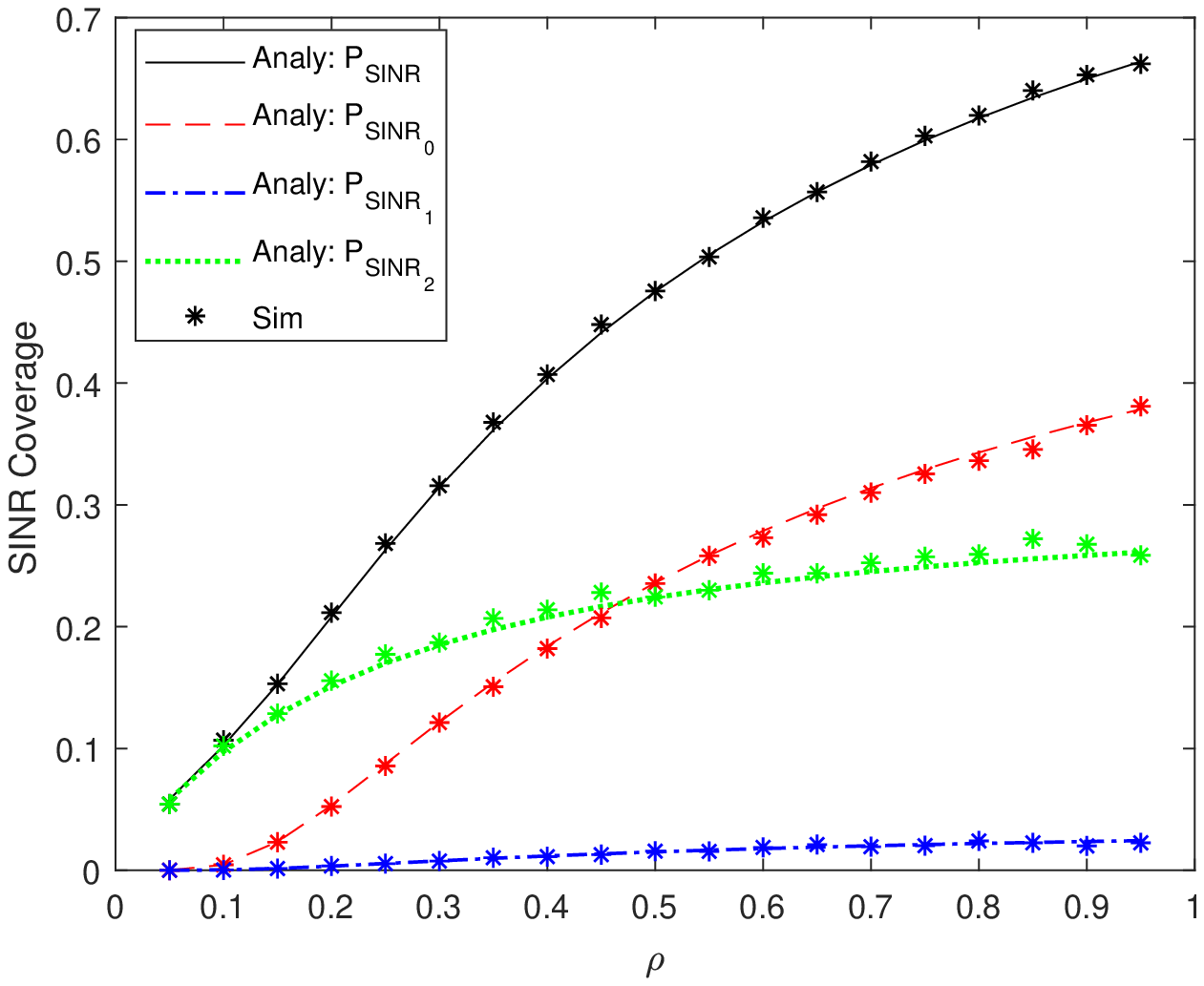}
		\subcaption{\scriptsize SINR coverage.}
	\end{minipage}
	\begin{minipage}{0.32\textwidth}
		\centering
		\includegraphics[width=1\textwidth]{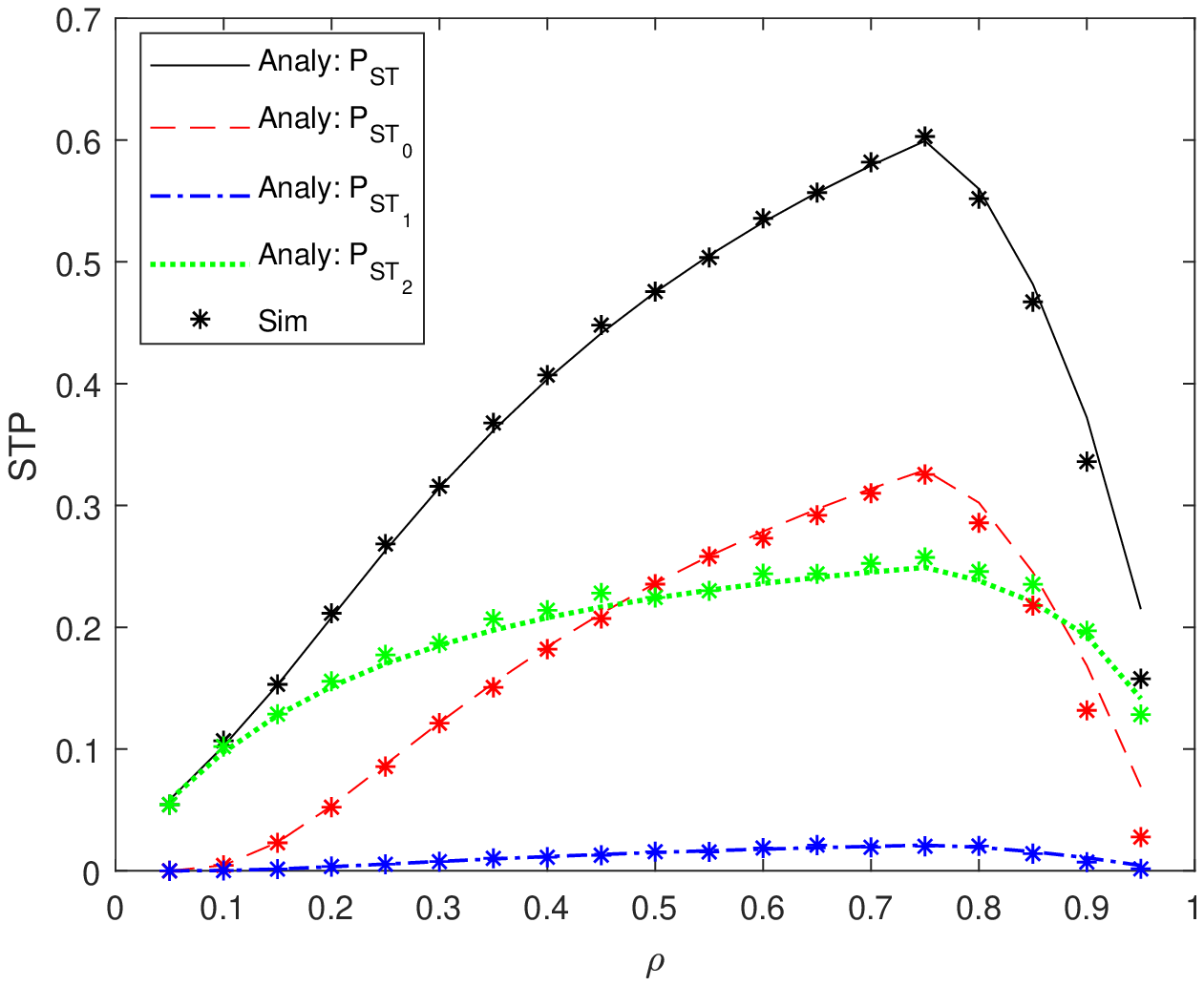}
		\subcaption{\scriptsize Successful transmission probability.}
	\end{minipage}
	\caption{\small STP, EC and SINRC as functions of the power splitting parameter $\rho$ when $\sigma=10$, $\tau=T$,  $\gamma_E=-40$ dB, $\gamma_{sinr}=-15$ dB. To show the impact of $\rho$, we use $\sigma_c=-10$ dB in this figure .  \normalsize}
	\label{figure_rho}
\end{figure}
In this subsection, we investigate the impact of the power splitting parameter $\rho$ on the network performance. From Fig. \ref{figure_rho}, we can conclude that EC is a monotonically decreasing function of $\rho$ and SINRC is an increasing function of $\rho$, due to the facts that larger $\rho$ means that more power is used for harvesting energy and less power for information decoding. Using the given set of parameter values, we observe that there exists an optimal $\rho$ value that maximizes the system downlink performance.
Since in this model, the interference is negligible, we can use (\ref{rho}) to approximately find the optimal value of $\rho$. By substituting the parameter values provided in  (\ref{rho}), we obtain $\rho = 0.7603$ and this is consistent with what we have from the numerical result.

\subsection{Impact of the $\tau$}
%\begin{figure}
%	\centering
%	\includegraphics[width=0.45\textwidth]{Uplink_rate2.eps}
%%	\includegraphics[width=0.45\textwidth]{Uplink_rate_joint2.eps} \\
%	\caption{\scriptsize Averaged uplink throughput as a function of $\tau$ for Thomas cluster process, when $\rho=0.5$, $\gamma_{sinr}=0$ dB, $\gamma^{UL}=-20$dB, and $\sigma=10$. }
%	\label{Uplink_rate}
%\end{figure}
\begin{figure}
	\centering
	\begin{minipage}{0.45\textwidth}
		\centering
		\includegraphics[width=1\textwidth]{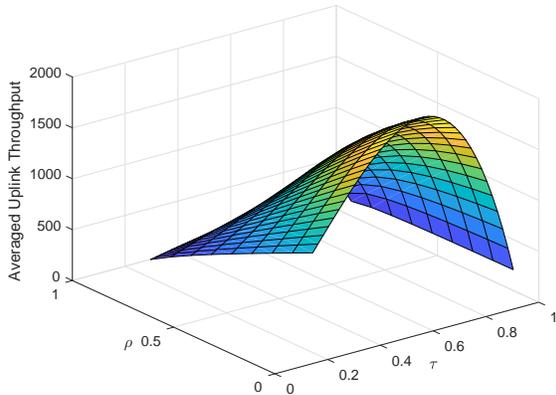}
		\subcaption{\scriptsize Averaged uplink throughput.}
	\end{minipage}
	\begin{minipage}{0.45\textwidth}
		\centering
		\includegraphics[width=1\textwidth]{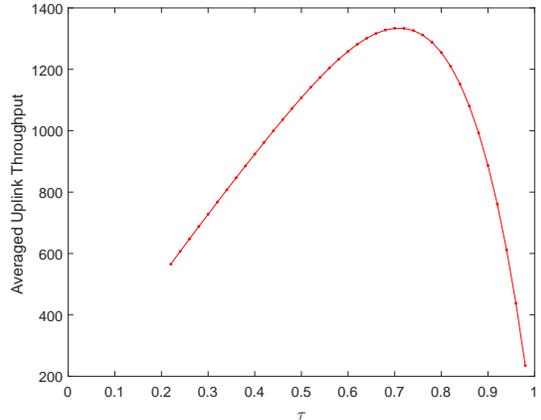}
		\subcaption{\scriptsize Averaged uplink throughput $\rho=0.5$.}
	\end{minipage}
	\caption{\small Averaged uplink throughput as a function of $\tau$ and $\rho$ for Thomas cluster process, when $\gamma_E = -40$ dB,  $\gamma_{sinr}=0$ dB, $\gamma^{UL}=-20$ dB, and $\sigma=10$. \normalsize}
	\label{Uplink_rate}
\end{figure}
In this section, we investigate the effect of  the time duration $\tau$ allocated to the downlink phase. Fig. \ref{Uplink_rate} shows the average uplink throughput as a function of $\tau$ and $\rho$ under the constraint that the average downlink throughput $R^{DL}$ is larger than $R_{\min}$ (addressing the optimization problem in (\ref{Rate_Opt})). As shown in Fig. \ref{Uplink_rate}(a), $R^{UL}$ decreases with increasing $\rho$, since larger $\rho$ means less power for energy harvesting. On the other hand, since $R^{DL}$ is a monotonically increasing function of $\tau$, if we want to satisfy the minimum throughput requirement, there is a minimum value of $\tau$. Therefore, as shown in Figs. \ref{Uplink_rate} (a) and (b), when $\tau$ is smaller than a certain value, the minimum downlink throughput constraint cannot be satisfied and the optimization problem in (\ref{Rate_Opt}) is not feasible. When $\tau$ increases, the downlink constraint is satisfied, and we note that there is an optimal $\tau$ that maximizes the average uplink throughput.

\section{Summary} \label{Con}
In this paper, we have jointly considered the downlink SWIPT and uplink information transmission in  UAV-assisted mmWave cellular networks, in which the UE locations are modeled using Thomas cluster processes and Mat\'ern cluster processes. Distinguishing features of mmWave communications, such as different path loss models for LOS and NLOS links, and directional transmissions, are taken into account.  We have characterized the CCDF and PDF of the distance from the typical UE to its own cluster center UAV, the nearest PPP-distributed UAV and the nearest GBS. In the downlink phase, we have determined the association probabilities of each tier BS. In addition, we have considered the power splitting technique in the SWIPT scenario, which allows the UEs to harvest energy and decode information simultaneously using the same received signal. We have characterized the  energy and SINR coverage probabilities of the considered UAV-assisted mmWave cellular network. Moreover, we have defined the successful transmission probability to jointly analyze the energy and SINR coverages and we have provided general expressions. In the uplink phase, we have considered the scenario that each UAV receives information from its own cluster member UEs. SINR coverage has been derived and general expressions are provided. In addition, we have formulated the average uplink throughput, aiming to find the optimal time division multiplexing for downlink and uplink phases. Finally, via numerical results we have investigated the impact of key system parameters on the network performance. We have shown that the system performance is improved when the cluster size becomes smaller. In addition, we have analyzed the optimal height of UAVs and optimal power splitting value that maximize the system performance. Optimal time division has also been addressed to maximize the average uplink throughput. We have verified that Thomas cluster processes and Mat\'ern cluster processes can lead to similar system performance trends.

\appendix
\subsection{Proof of Lemma 1 }  \label{Proof_0TH_CCDF}
We can express the CCDF as
\begin{align}
\Fbar_{R^s_{0}} (x) &=\prob(r>x|\text{the link can be } s)  \notag
  = \frac{\prob(r>x,\text{the link is } s )}{\prob(\text{the link can be } s)} \notag\\
&  = \frac{\E_D[\prob(r>x|s,D)\prob(s|D)]}{D^s_0} \notag
  = (D^s_0)^{-1} \E_D[\prob(\sqrt{D^2+H^2}>x|s,D)p^s_j(\sqrt{D^2+H^2})] \notag \\
&  = (D^s_0)^{-1} \E_D[\prob(D>\sqrt{x^2-H^2})p^s_j(\sqrt{D^2+H^2})] \notag \\
&  = (D^s_0)^{-1} \int_{\sqrt{x^2-H^2}}^{\infty} p^s_j(\sqrt{d^2+H^2}) f_D(d) \, \dsp d
\end{align}
where $x\geq H$, $s\in\{\los, \nlos\}$, and $D^s_0=\int_{0}^{\infty} p^s_U(r) f_{R_0} (r)dr$ is the probability that the link between the typical UE and its cluster center UAV is in $s$ transmission.

Therefore, we can get the PDF as follows:
\begin{align}
 f_{R^s_{0}} (x) &= - \frac{d \Fbar_{R^s_{0}} (x)  }{dx} \notag
 =\frac{x}{\sqrt{x^2-H^2}} p^s_U(x) f_D(\sqrt{x^2-H^2})  / D^s_0.
%=\frac{x p^s_U(x)}{{\sigma^2_j D^s_0}}\exp\left(\frac{H^2-x^2)}{2{\sigma^2}}\right)
\end{align}
%where $x\geq H$.

\subsection{Proof of Lemma 2} \label{Proof_UAV_CCDF}
Since the UAVs are assumed to be located at the same hight, the distribution of the UAVs is a 2-D PPP. Given that there is at least one LOS/NLOS UAV around the UE, we have
\begin{align}
\Fbar_{R^s_U}(x) &= \prob(r>x|\text{there has at least one $s$ UAV around})\notag \\
&      =\frac{\prob(r>x, \text{the link is in $s$ transmission})} {\prob(\text{there has at least one $s$ UAV around} )} \notag \\
&      =\frac{\prob(\text{there is no $s$ UAV loser than x})}{D^s_U} \notag \\
&      \overset{(a)}{=} (D^s_U)^{-1} e^{- \lambda^s_U A(x)} \notag
      \overset{(b)}{=} (D^s_U)^{-1} e^{- \int \int \lambda_U p^s_U(r) t dt  d\theta}  \notag \\
&      = (D^s_U)^{-1} e^{- 2 \pi  \lambda_U \int_0^{\sqrt{x^2-H^2}} p^s_U(\sqrt{t^2 +H^2}) t dt}  \notag \\
&      = (D^s_U)^{-1} e^{- 2 \pi  \lambda_U \int_H^{x} p^s_U(t) t dt}
\end{align}
where $s\in\{\los, \nlos\}$, $A(x)$ is the area of the circle with radius $x$ in the air, $\lambda^s_U$ is the density of $s$ UAVs, (a) is from \cite[Section III. A]{Cellular_JAndrews} and (b) follows from the integration of the area using polar coordinates.

\subsection{Proof of Lemma 3} \label{Proof_AP}
Let us define two events $S_1=\{$The typical UE is associated with a $j^{th}$ tier BS$\}$ and $S_2=\{$The associated link is in  $s\in\{\los,\nlos\}$ transmission$\}$. Now we have
\begin{align}
A_{j,s} &\overset{(a)}{=} \prob (S_1 S_2)
\overset{(b)}{=}\prob(S_2) \prob(S_1 |S_2)
\end{align}
%\begin{align}
%	 A_{j,s} &\overset{(a)}{=} \prob (S_1 S_2)
%	 \overset{(b)}{=}  \prob(S_2) \prob(S_1 |S_2) \notag \\
%	& =\prob(L_{j,s}^{-1}>L_{j,s'}^{-1}) \prob(P_jB_jL_{j,s}^{-1}>P_kB_kL_k^{-1}, k\in\{0,1,2\},k\neq j)\notag \\
%	& =  \prob( L_{j,s'}> L_{j,s}) \prob(L_k>\frac{P_k B_k}{P_j B_j} L_{j,s}, k\in \{0,1,2\}, k\neq j)
%\end{align}
where (a) is due to the definition of the association probability, and (b) follows from the Bayes' theorem.
\subsubsection{The association probability of the $0^{th}$ tier UAV}
\begin{align}
			A_{0,s} \notag
			& \overset{(a)}{=} D^s_0 \prob(P_0 B_0L_{0,s}^{-1}>P_kB_kL_k^{-1}, k\in\{1,2\})
             \overset{}{=} D^s_0 \prob\left(L_k>\frac{P_k B_k}{P_0 B_0} L_{0,s}, k\in \{1,2\}\right)    \notag \\
%			& \overset{(a)}{=} \prob (\text{ the main link is in } s\in \{\los, \nlos \})
%			\prob(L_{k,L}>C_{k0} L_{0,s} \text{ and }L_{k,N}>C_{k0} L_{0,s} ,k\in\{1,2\})       \notag \\
			&= \E_{R^s_{0}} \Big[  D_0^s \prod_{k} \prod_{b} D_k^b \prob(\kappa_k^b r_k^{\alpha_k^b } > C_{k0} \kappa_0^{s} r_0^{\alpha^{s}_0})\Big]
			 = \E_{R^s_{0}} \left[  D_0^s \prod\limits_{k} \prod\limits_{b} D_k^b  \Fbar_{R_k^b} \left(Q_{k0}^{sb} (r_0)\right)\right]
\end{align}
where  $C_{kj}=\frac{P_k B_k}{P_j B_j} $, $Q_{kj}^{sb}(r)=\left(\frac{P_k B_k \kappa_j^s}{P_j B_j \kappa_k^b} r^{\alpha^s_j}\right)^{\frac{1}{\alpha^b_k}}$, $D^s_0=\int_{0}^{\infty} p^s_U(r) f_{R_0} (r)dr$ is the probability that the link from the typical UE to its own cluster center UAV can be in $s$ transmission, and (a) is due to the fact that there is only one UAV in the $0^{th}$ tier and the LOS BSs and NLOS BSs in the $1^{st}$ and $2^{nd}$ tier are independent.
\subsubsection{The association probability of the $j^{th}$ tier BSs}
\begin{align}
			 &A_{j,s}  =\prob(L_{j,s}^{-1}>L_{j,s'}^{-1}) \prob(P_jB_jL_{j,s}^{-1}>P_kB_kL_k^{-1}, k\in\{0,1,2\},k\neq j)\notag \\
			 & =  \prob( L_{j,s'}> L_{j,s}) \prob(L_k>\frac{P_k B_k}{P_j B_j} L_{j,s}, k\in \{0,1,2\}, k\neq j)  \notag \\
%			  &\overset{}{=} \prob( L_{j,s'}> L_{j,s}) \prob(L_0>C_{0j} L_{j,s})\prob(L_k>C_{kj} L_{j,s}) \notag \\
			&= \prob( L_{j,s'}> L_{j,s})  \prob(L_{0,L}>C_{0j} L_{j,s}  \text{ or } L_{0,N}>C_{0j} L_{j,s} )
			 \prob(L_{k,L}>C_{kj} L_{j,s} \text{ and }L_{k,N}>C_{kj} L_{j,s} ,k\neq j)       \notag \\
			& = D_j^s \E_{R^s_{j}} \Bigg[D_j^{s'} \prob(\kappa_j^{s'} r_j^{\alpha_j^{s'}}>\kappa_j^{s} r_j^{\alpha_j^{s}})
			 \left(\sum_b D_0^b  \prob\left(\kappa_0^b r_0^{\alpha_0^b } > C_{0j} \kappa_j^{s} r_j^{\alpha^{s}_j}\right)\right)  \prod_{b} D_k^b \prob(\kappa_k^b r_k^{\alpha_k^b } > C_{kj} \kappa_j^{s} r_j^{\alpha^{s}_j})\Bigg] \notag \\
			& =  D_j^s\E_{R^s_{j}}  \Bigg[D_j^{s'} \Fbar_{R^{s'}_{j}}\left(Q_{jj}^{ss'} (r_j)\right)
			 \left(\sum\limits_b D_0^b \Fbar_{R^b_{0}}\left(Q_{0j}^{sb}(r_j) \right)\right) \prod\limits_b D_k^b  \Fbar_{R_k^b} \left(Q_{kj}^{sb} (r_j)\right) \Bigg]
\end{align}
where $s,s' \in \{\los,\nlos\}$, $s' \neq s$, $D^s_j=1-e^{-2 \pi \lambda_j \int_0^\infty t p^s_j(t) dt}$ is the probability that the typical UE has at least one LOS/NLOS $j^{th}$ tier BS around.

\subsection{Proof of Theorem 1 } \label{Proof_EC}
Given $S_{j,s} =$ \{The typical UE is associated with a LOS/NLOS BS in the $j^{th}$ tier\}, we can express the conditional energy coverage probability as
\begin{align}\label{Proof_EC_total}
		 P^c_{E_{j,s}} (\rho,\tau, \gamma_E)
		    &\overset{(a)}{=}\prob(\tau(1-\rho)(P_m +I >\gamma_E | S_{j,s})
	  		\overset{(b)}{=} \sum_{n=0}^N (-1)^n \Big(\substack{N \\ \\n}\Big) \E \left[ e^{-\ahat (P_m +I ) } \right] \notag \\
	  		&\overset{(c)}{=} \sum_{n=0}^N (-1)^n \Big(\substack{N \\ \\n}\Big) \E \left[ \E_{h_0} \left[ e^{-\ahat P_m} \right] \prod_k \E \left[e^{-\ahat I_k} \right] \right]  \notag\\
	  		&\overset{(d)}{=} \sum_{n=0}^N (-1)^n \Big(\substack{N \\ \\n}\Big)  \E \bigg[ % \left(1+\frac{\ahat P_j G }{\kappa^s_j r^{\alpha^s_j} N_s}\right)^{-N_s} \notag \\
	  		\left(1+\ahat P_j G (\kappa^s_j r^{\alpha^s_j} N_s)^{-1}\right)^{-N_s}
	  		 \prod_k \cL_{I_k}(\ahat) \bigg]
%	  		&\hspace{0.9in} \E_{ I_1| P_j, S_{j,s}} \left[e^{-\ahat I_1} \right]  \E_{ I_0 | P_j, S_{j,s}} \left[e^{-\ahat I_0} \right] \bigg]
\end{align}
where $P_m$ is the received power of the main link,  $I=I_0+I_1+I_2$ is the total  interference, $\ahat=\frac{an\tau(1-\rho)}{T}$, $a=N(N!)^{-\frac{1}{N}}$. (a) follows from the definition of energy coverage. (b) is modified from \cite[Appendix A]{EH_TKhan}. (c) is due to the independence of $P_m$, $I_0$, $I_1$ and $I_2$ given $S_{j,s}$. (d) is  calculated by using the moment generating function (MGF) of a normalized Gamma random variable.

\subsubsection{For $I_0$}
Since $I_0$ only exists when the typical UE is associated with the $1^{th}$ tier UAVs or the $2^{nd}$ tier GBSs, we have
\begin{align}\label{Proof_EC0}
%		\E_{ I_0 | P_m,S_{j,s}}[e^{-\ahat I_0} ]
        \cL_{I_0}(\ahat)
			&\overset{(a)}{=} \sum\limits_{G} \sum\limits_{b} p_G D^b_0 \cL_{I^{bG}_0} (\ahat) \\
			 %\E_{ r_0 | P_m,S_{j,s}}
		\cL_{I^{bG}_0} (\ahat)
		    &=\E_{r_0|S_{j,s}}\left[\E_{h_0} \left[\exp\left(-\ahat P_0 G h_0 \left(\kappa^b_0  r_0^{\alpha^b_0}\right)^{-1} \right) \right] \right] \notag \\
			&\overset{(b)}{=} \E_{r_0|S_{j,s}} \left[	\left(1+\ahat P_0 G (\kappa^b_0 r_0^{\alpha^b_0} N_b)^{-1}\right)^{-N_b} \right] \notag\\
			&= \int_{max(H,Q_{0j}^{sb}(r))}^\infty \frac{f_{R^b_{0}}\left(r_0|r_0>Q_{0j}^{sb}(r) \right)}{\left(1+\ahat P_0 G (\kappa^b_0 r_0^{\alpha^b_0} N_b)^{-1}\right)^{N_b}}
			  dr_0 \notag \\
			&=  \int_{max(H,Q_{0j}^{sb}(r))}^\infty \frac{ f_{R^b_{0}}(r_0) dr_0}{\left(1+\ahat P_0 G (\kappa^b_0 r_0^{\alpha^b_0} N_b)^{-1}\right)^{N_b} D^b_0\Fbar_{R^b_{0}}(Q_{0j}^{sb}(r))}
\end{align}
where $s, b\in \{\los,\nlos\}$, $G\in \{M_bM_u,M_bm_u,m_bM_u,m_bm_u,\}$, $p_G$ is the probability for $G$, (a) is due to the fact that there is only one BS in the $0^{th}$ tier  which can be in LOS or NLOS transmission with antenna gain $G$, and (b) is because of the MGF of a normalized Gamma random variable.

\subsubsection{For $I_k$ ($k=1,2$)}
%From \cite[Appendix A]{EH_TKhan}, we can get
\begin{align}
		\cL_{I_k}(\ahat) &= \prod_G \prod_b \cL_{I^{bG}_k} (\ahat) \\
		\cL_{I^{bG}_0} (\ahat)&	=   E_{r_k|S_{j,s}} \left[ \left(1+\ahat P_k G (\kappa^b_k r_k^{\alpha^b_k} N_b)^{-1}\right)^{-N_b} \right] \notag \\
		&\overset{(a)}{=} e^{-2 \pi \lambda_k p_G \int_{B_{d2}}^\infty \left(1-\left(1+\ahat P_k G (\kappa^b_k r_k^{\alpha^b_k} N_b)^{-1}\right)^{-N_b} \right) p^b_k(r_k) r_k dr_k } \label{Proof_EC1}
\end{align}
where $B_{d2}=\max(H,Q^{sb}_{kj}(r))$ for $k=1$, and $B_{d2}=Q^{sb}_{kj}$ for $k=2$. (a) follows by computing the moment generating functional (MGFL) of PPP.

Therefore, by substituting (\ref{Proof_EC0}) - (\ref{Proof_EC1}) into (\ref{Proof_EC_total}), we can obtain (\ref{EC}).

\subsection{Proof of Theorem 2} \label{Proof_STP}
Given $S_{j,s}$, we can express the conditional successful transmission probability as
\begin{align}
		 &P^c_{\suc_{j,s}} (\rho, \tau,\gamma_E, \gamma_{sinr}) \notag \\
			& \overset{(a)}{=} \prob\left(\tau(1-\rho) (P_m+I) >\gamma_E, \frac{P_m}{\frac{\sigma_c^2}{\rho}+\sigma_n^2+I}>\gamma_{sinr} \Big| S_{j,s} \right) \notag \\
			&=\prob\left(P_m > T_1, P_m >T_2 \Big| S_{j,s} \right) \notag \\
			& \overset{ }{=} \prob\left(P_m > T_1, \Big| S_{j,s} \right) \indictor \left(T_1> T_2 \Big| S_{j,s}  \right)
			+ \prob\left(P_m > T_2 \Big| S_{j,s} \right)  \indictor\left(T_1 \leq T_2 \Big| S_{j,s}  \right) \notag\\
			&  \overset{(b)}{=}\E_{I} \left[\prob\left(P_m >T_1, \Big| S_{j,s} \right) \right] \E_{I} \left[ \indictor \left( T_1> T_2 \Big| S_{j,s}  \right)  \right] +
			 \E_{I} \left[	 \prob\left(P_m > T_2, \Big| S_{j,s} \right) \right] \E_{I} \left[ \indictor\left( T_1 \leq T_2 \Big| S_{j,s}  \right)	\right] \notag\\
			& \overset{(c)}{=}P^c_{E_{j,s}} (\rho, \gamma_E) \E_{I} \left[ \indictor \left(I<\omega  \Big| S_{j,s}  \right) \right] +
			 P^c_{SINR_{j,s}} (\rho, \gamma_{sinr}) \E_{I} \left[\indictor \left(I \geq \omega \Big| S_{j,s}  \right) \right] \notag\\
			& \overset{}{=}P^c_{E_{j,s}} (\rho, \gamma_E)(1-\Fhat_{I} \left(\omega \right) ) +P^c_{SINR_{j,s}} (\rho, \gamma_{sinr})  \Fhat_{I} \left(\omega \right)
\end{align}
where $T_1= \frac{\gamma_E}{\tau(1-\rho)}-I $, $T_2= \gamma_{sinr} \left( \frac{\sigma_c^2}{\rho}+\sigma_n^2 +I \right)$, $\omega=\frac{1}{1+\gamma_{sinr}} \left( \frac{\gamma_E}{\tau(1-\rho)}-\gamma_{sinr}\left(\frac{\sigma_c^2}{\rho}+\sigma_n^2\right)\right)$, $\Fhat_I(x)$ is the CCDF of $I$, (a) follows from the definition of the successful transmission probability, (b) is due to the fact that given $I$, the indicator function is a constant and is independent to $P_m$, (c) follows from the definition of energy and SINR coverage probability and $\omega$ is obtained by $ \frac{\gamma_E}{\tau(1-\rho)}-I>\gamma_{sinr} \left( \frac{\sigma_c^2}{\rho}+\sigma_n^2 +I\right) $.

Since $\Fhat_{I}(x)=\prob(I>x)$ and this is similar as the energy coverage probability when $P_m=0$, we can obtain the expression of $\Fhat_{I}(x)$ following the derivation in Appendix \ref{Proof_EC}  by replacing $\gamma_E$ and $\rho$ with $x$ and 0.

\subsection{Proof of Theorem 3} \label{Proof_UplinkSINR}
We can express the SINR coverage probability in the uplink phase as
\begin{align}
		P^{UL}_{SINR} (\gamma_{UL})%&=\prob \left(SINR^{UL} \geq \gamma……{UL} \right) \notag \\
		        &= \prob \left(\frac{P^{UL}_t G_0 h_0 (k_U^s r_0^{\alpha^s_U})^{-1}}{\sigma_n^2 + I_{user}}  \geq \gamma^{UL} \right) \notag\\
		        &=\prob \left(h_0 \geq \frac{\gamma^{UL}r_0^{\alpha^s_U} }{P^{UL}_t G_0 k_U^s } (\sigma_n^2 +I_{user}) \right) \notag \\
		        &\overset{(a)}{=} \E_{R^s_0} \left[ \sum_{n=0}^{N_s} (-1)^{n+1} \binom{N_s}{n} \exp\left( - \frac{n \eta_s \gamma^{UL}r_0^{\alpha^s_U} }{P^{UL}_t G_0 k_U^s } (\sigma_n^2 +I^L_{user}+ I^N_{user}) \right) \right] \notag \\
		       % &= \E \left[ \sum_{n=0}^{N_s} (-1)^{n+1} \binom{N_s}{n} \exp\left( - \mu^{UL}_s (\sigma_n^2 +I_{user}) \right) \right] \notag \\
		        &=\E_{R^s_0} \left[ \sum_{n=0}^{N_s} (-1)^{n+1} \binom{N_s}{n} e^{ - \mu^{UL}_s \sigma_n^2 } \cL_{ I^L_{user} }( \mu^{UL}_s) \cL_{ I^N_{user} }( \mu^{UL}_s) \right]
\end{align}
where $\mu^{UL}_s=\frac{n \eta_s \gamma^{UL}r_0^{\alpha^s_U} }{P^{UL}_t G_0 k_U^s }$ and (a) follows by computing the MGF of the gamma random variable $h_0$. $\cL_{ I^b_{user} }( \mu^{UL}_s)$ is the Laplace transform expression and can be further analyzed as follows:
\begin{align}
\cL_{ I^b_{user} }( \mu^{UL}_s)=\prod_G \cL_{ I^{bG}_{user} }( \mu^{UL}_s)
\end{align}
where
\begin{align}
&\cL_{ I^{bG}_{user} }( \mu^{UL}_s) = \E \left[\exp \left( -\mu^{UL}_s I^{bG}_{user}   \right) \right] \notag \\
&= \E \left[ \exp \left( -\mu^{UL}_s \sum\limits_{i\in \Phi^b_{user}} P^{UL}_t G_i h_i (\kappa^b_U r_{i}^{\alpha^b_U})^{-1}   \right)\right]  \notag \\
&\overset{}{=} \E_{\bf{x}} \left[ \prod_{\bf{x}\in\Phi_U} \E_{\bf{y}} \left[ \E_{h_i} \left[ \exp \left( -\mu^{UL}_s  P^{UL}_t G_i h_i (\kappa^b_U (|| \xbf + \ybf||^2+H^2)^{\frac{\alpha^b_U}{2}})^{-1}   \right)\right] \right] \right]  \notag \\
&\overset{(a)}{=} \E_{\bf{x}} \left[ \prod_{\bf{x}\in\Phi_U} \E_{\bf{y}} \left[  \frac{1}{ \left( 1+\mu^{UL}_s  P^{UL}_t G_i (\kappa^b_U (|| \xbf + \ybf||^2+H^2)^{\frac{\alpha^b_U}{2}} N_b)^{-1} \right)^{N_b} }   \right] \right]  \notag \\
&\overset{(b)}{=} \E_{\bf{x}}  \left[ \prod_{\bf{x}\in\Phi_U} \iint_{R^2}  \frac{1}{ \left( 1+\mu^{UL}_s  P^{UL}_t G_i (\kappa^b_U (|| \xbf + \ybf||^2+H^2)^{\frac{\alpha^b_U}{2}} N_b)^{-1} \right)^{N_b} } f_{\bf{Y}}(y)dy \right] \notag \\
&\overset{(c)}{=} \exp\left(-p_G\lambda^b_{user} \iint_{R^2}\left(  1- \iint_{R^2}\frac{1}{\left( 1+\mu^{UL}_s  P^{UL}_t G_i (\kappa^b_U (|| \xbf + \ybf||^2+H^2)^{\frac{\alpha^b_U}{2}} N_b)^{-1} \right)^{N_b} } f_{\bf{Y}}(y)dy \right) dx \right) \notag \\
& \overset{(d)}{=} \exp\left(-2 \pi p_G \lambda^b_{user} \int_{0}^{\infty}\left(  1- \int_{0}^{\infty} \left( 1+\mu^{UL}_s  P^{UL}_t G_i (\kappa^b_U (v^2+H^2)^{\frac{\alpha^b_U}{2}} N_b)^{-1} \right)^{-N_b}  f(v|w)dv \right) wdw \right)
%&\overset{}{=} \exp\left(-2 \pi\lambda^b_{user} \int_{0}^{\infty}\int_{0}^{\infty}\frac{1}{ 1+(\mu^{UL}_s  P^{UL}_t G)^{-1} \kappa^b_U (v^2+H^2)^{\frac{\alpha^b_U}{2}} } f(v|w)dv wdw \right) \notag \\
%&\overset{e}{=} \exp\left(-2 \pi\lambda_U \int_{0}^{\infty}\frac{p_{active}p_G p^b_U(\sqrt{v^2+H^2})}{ 1+(\mu^{UL}_s  P^{UL}_t G)^{-1} \kappa^b_U (v^2+H^2)^{\frac{\alpha^b_U}{2}} } vdv \right)
\end{align}
where $\bf{y}$ is the coordinate of the interfering UE with respect to the projection of its cluster center, $\bf{x}$ is the coordinate of the ground projection of that cluster center with respect to the ground projection of the typical UAV, and $w=||\bf{x}||$ and $v=||\bf{x+y}||$. (a) is due to the MGF of $h_i$. (b) follows from the definition of expectation. (c) is due to the computation of the probability generating functional (PGFL) of PPP. (d) is obtained by converting the coordinates from Cartesian to polar. %(e) follows from an approximation base on the Rician distribution property that $\int_{0}^{\infty} f(v|w) w dw =v$.

\end{spacing}

\begin{spacing}{1}
\bibliographystyle{IEEEtran}
\bibliography{cluster_milli}
\end{spacing}

\end{document}